\documentclass[twocolumn,]{aastex631}

\shorttitle{Landscape of Unstable Mass Transfer}
\shortauthors{Twum et al.}
\usepackage{xcolor} 
\usepackage{multirow}
\usepackage{amsmath}
\usepackage{graphicx}
\usepackage{afterpage}

\usepackage{xcolor}
\definecolor{ochre}{rgb}{0.8, 0.47, 0.13}

\definecolor{green}{rgb}{0.1, 0.5, 0.1}
\newcommand{\aat}[1]{{\color{green}#1}}

\begin{document}

\title{The Landscape of Unstable Mass Transfer in Interacting Binaries and Its Imprint on the Population of Luminous Red Novae}  

\correspondingauthor{Angela A. G. Twum}
\email{atwum@ucsc.edu}

\author[0009-0006-4675-7596]{Angela A. G. Twum}
\affiliation{Department of Astronomy and Astrophysics, University of California, Santa Cruz, CA 95064, USA}

\author[0000-0003-1817-3586]{Alejandro Vigna-G\'{o}mez}
\affiliation{Niels Bohr International Academy, Niels Bohr Institute, University of Copenhagen, Blegdamsvej 17, 2100, Copenhagen, Denmark}

\author[0000-0002-1417-8024]{Morgan~MacLeod} 
\affiliation{Center for Astrophysics, Harvard \& Smithsonian, 60 Garden Street, MS-16, Cambridge, MA 02138, USA}
\affiliation{Institute for Theory and Computation, Harvard \& Smithsonian, 60 Garden Street, MS-51, Cambridge, MA 02138, USA}

\author[0000-0003-0381-1039]{Rosa Wallace Everson}
\affiliation{Department of Astronomy and Astrophysics, University of California, Santa Cruz, CA 95064, USA}

\author[0000-0003-0381-1039]{Ricardo~Yarza} 
\altaffiliation{NASA FINESST Fellow}
\affiliation{Department of Astronomy and Astrophysics, University of California, Santa Cruz, CA 95064, USA}

\author[0000-0002-5814-4061]{V. Ashley Villar} 
\affiliation{Center for Astrophysics, Harvard \& Smithsonian, 60 Garden Street, MS-16, Cambridge, MA 02138, USA}
\affiliation{The NSF AI Institute for Artificial Intelligence and Fundamental Interactions}

\author[0000-0003-2558-3102]{Enrico~Ramirez-Ruiz} 
\affiliation{Department of Astronomy and Astrophysics, University of California, Santa Cruz, CA 95064, USA}

\begin{abstract}
A common-envelope (CE) phase occurs when a star engulfs its companion and is widely considered the primary channel for producing Luminous Red Novae (LRNe). 
In this study, we combine binary-population synthesis with stellar-evolution calculations to systematically estimate the mass, velocity, and launching radius of ejecta produced during coalescence across a range of binary configurations. 
Our aim is to quantify how unstable mass-transfer dynamics in binaries at various evolutionary stages shape CE outcomes, enabling a predictive framework for modeling the LRN luminosity function.
We find a bimodal distribution of plateau luminosities with significant implications for binary mass stability criteria that can be tested with forthcoming LSST observations.
This bimodality emerges from differing mass-ejection outcomes during common-envelope interactions, which can lead either to stellar mergers, often accompanied by tidal disruption of the companion, or to successful envelope ejection. 
Although our predicted plateau luminosities and timescales broadly match existing observations, the models underpredict the number of LRNe with long-duration plateaus ($t_p \gtrsim 100\, \text{d}$) by about a third. We propose that these long-duration events arise from highly extended progenitors whose envelopes are ejected over multiple orbits (i.e., non-impulsively), producing relatively faint, long-lived transients.
By constraining ejecta properties and incorporating pre-outburst progenitor imaging, we show how our models can clarify the physical processes that drive unstable mass transfer in these events.
Finally, we argue that common-envelope interactions involving white-dwarf accretors can yield exotic outcomes, including red giants containing embedded white dwarfs that resemble Thorne--\.{Z}ytk{\'o}w objects (TŻOs), along with calcium-rich supernovae that preserve hydrogen envelopes.
\end{abstract}

\section{Introduction} 
\label{sec:intro}
More than half of the stars in our Solar neighborhood are members of binary systems \citep[e.g.,][]{Sana2012}. Interacting binaries offer a unique laboratory for studying stellar-evolution processes that cannot be probed in isolated stars \citep[e.g.,][]{2020ApJ...901...44W,2024ApJ...977...16R}. Understanding the formation and evolution of binary stars is essential for probing extreme astrophysical processes that occur under high-energy and high-density conditions \citep[e.g.,][]{1996A&A...309..179P,apjac6269bib12,apjac6269bib106,apjac6269bib19,apjac6269bib18,2024ApJ...971..132E}. These systems, the progenitors of compact binaries that produce gravitational waves \citep[e.g.,][]{apjac6269bib111,apjac6269bib5,Vigna-Gomez2018}, are essential for comprehending diverse transient phenomena, including type Ia supernovae \citep[e.g.,][]{1973ApJ...186.1007W,2014A&A...563A..83C} and gamma-ray bursts \citep[e.g.,][]{2004MNRAS.348.1215I,apjac6269bib88,2019ApJ...883L..45F,2022A&A...657L...8B}.

As stars evolve, their hydrogen-rich outer layers expand significantly during the later stages of their lives. In close binary systems, this expansion can lead to mass exchange when material that was initially gravitationally bound to one star becomes susceptible to the potential of its companion.

The effective potential in a binary system is commonly approximated using three components: the gravitational potentials of the two stars and the rotational potential of the co-rotating frame, which depends on the binary separation \citep[e.g.,][]{apjac6269bib84,apjac6269bib85,apjac6269bib51}. The resulting equipotential surfaces define regions where the gravitational influence of each star dominates. The critical surface, known as the Roche equipotential surface, encloses the two Roche lobes associated with the stars \citep[e.g.,][]{apjac6269bib71,2023MNRAS.524..471C}.

When a star expands beyond the boundary of its Roche lobe, material can flow toward the companion through the inner Lagrange point, forming a \textit{nozzle} between the two stars \citep{1975ApJ...198..383L,apjac6269bib71}. As the transferred material carries angular momentum, it modifies the orbital parameters of the binary system \citep{1981ARA&A..19..277S}. Additionally, changes in the binary mass ratio further alter the orbit, even if the total angular momentum of the system remains constant \citep{1979ApJ...229..223S}.

Once mass transfer begins in a binary system, the stars may either merge or the system may survive the episode and remain a bound binary capable of further interaction \citep[e.g.,][]{2001ASPC..229..239P}. 

The outcome of a mass transfer episode is sensitive to the dynamical stability of the binary and its response to mass exchange.
This dynamical stability depends on several factors, including the mass ratio of the stars, the mechanisms for angular momentum loss, and the internal structure of the donor star \citep[e.g.,][]{apjac6269bib71,2025arXiv250521264S}. However, even with stable mass transfer, the survival of the binary is not guaranteed \citep[e.g.,][]{2002MNRAS.329..897H}. The characteristics of the surviving binary population are critically dependent on the interplay of these physical processes and can offer valuable insights into stellar evolution.

When mass transfer becomes unstable, runaway mass flow ensues, potentially resulting in a stellar merger \citep[e.g.,][]{apjac6269bib101, 2001ASPC..229..239P,apjac6269bib78,2018ApJ...863....5M,apjac6269bib38}. Under certain conditions, runaway mass transfer leads to the formation of a ``common envelope'' (CE) encompassing both stars in the binary \citep{apjac6269bib71,apjac6269bib110,apjac6269bib30,2013A&ARv..21...59I,apjac6269bib93}. If the companion star moves through the common envelope, its orbital motion slows due to frictional forces \citep[e.g.,][]{2015ApJ...803...41M}. As a result, the orbital separation between the two stars decreases significantly, and the potential energy released by the binary can be used to expel part of the common envelope \citep[e.g.,][]{2015ApJ...798L..19M,apjac6269bib44,apjac6269bib33,apjac6269bib47,2025ApJ...979L..11E}.

Contact-binary and common-envelope mergers are two pathways leading to the coalescence of stars. In contact binaries, the stars share a rotating envelope, while in common-envelope systems, one star is engulfed by its companion’s outer layers. In the latter, the embedded star may merge with its companion if the entire envelope is not expelled. For example, when the embedded object is a neutron star, this process has been predicted to result in the formation of a Thorne--\.{Z}ytkow object \citep[e.g.,][]{1975ApJ...199L..19T,1992ApJ...386..206C,2024ApJ...971..132E,2024ApJ...977..196H}. These mergers, along with the material expelled during the common-envelope phase, are thought to produce Luminous Red Novae (LRNe), optical/infrared transients that typically last weeks to months \citep[e.g.,][]{apjac6269bib55,apjac6269bib65,apjac6269bib9,apjac6269bib45,apjac6269bib114,apjac6269bib46,apjac6269bib100,apjac6269bib7,apjac6269bib10,apjac6269bib75,apjac6269bib107,apjac6269bib6}.

The light curves of LRNe provide valuable information about the properties of the merger ejecta \citep{apjac6269bib106}, particularly their mass and velocity \citep[e.g.,][]{2013Sci...339..433I}. By accurately inferring these properties, and combining them with insights from pre-outburst imaging of the progenitor, we can better understand the physical processes driving unstable mass transfer in these events \citep{Macleod2017,apjac6269bib6}. Consequently, LRNe may prove essential for understanding the nature of the unstable mass‑transfer processes operating in binary systems.

The coming decade is expected to be a pivotal period for discovering LRNe, particularly due to the large, uniformly selected samples enabled by extensive wide-field surveys \citep{Howitt2020}. The large number of expected observations, estimated at roughly 20–750 per year \citep{Howitt2020}, will enable statistical studies expected to shed light on binary evolution, particularly the physics of mass transfer and mergers. In this work, we integrate population synthesis with stellar evolution models to self-consistently determine the ejecta mass, velocity, and launching radius for specified binary configurations, and use the results to predict LRN observables such as light-curve duration and plateau luminosity. In Section~\ref{sec:compas}, we introduce the population synthesis formalism, initial model conditions, and stellar evolution protocol. The outline of the manuscript is as follows. Section~\ref{sec:donors} examines systems undergoing unstable mass transfer and common-envelope evolution, focusing on the donor and accretor states. 
In Section~\ref{sec:mesa}, we present a revision to the standard CE energy formalism which incorporates the effects of tidal disruption and present the resulting binary outcomes from stellar-structure calculations. 
Section~\ref{sec:landscape} derives LRN plateau durations and luminosities, and Section~\ref{sec:dis} discusses insights in the context of current observations.

\section{Population synthesis and Stellar evolution modeling}\label{sec:compas}

\subsection{Binary Populations with COMPAS}
Our goal is to characterize the progenitor populations of LRN transients and assess how uncertainties in initial conditions affect predictions for the frequency and distribution of binaries undergoing dynamically unstable mass transfer. To this end, we employ the binary population synthesis code COMPAS v02.49.05 \citep{Setvenson2017,Vigna-Gomez2018,TeamCompas2022,TeamCompas2025}. COMPAS uses simplified prescriptions for stellar and binary evolution to simulate evolution of binary-star systems from inception. The binary evolution model used in COMPAS is largely analogous to the binary stellar evolution population synthesis implementation presented in \citet{2002MNRAS.329..897H}, which enables the rapid simulation of binary-star systems, generating vast populations for robust statistical analyses.
COMPAS derives the properties of a star such as luminosity, radius, and core mass based on its  mass, metallicity, and age using analytical formulas that are fitted to detailed stellar models. The zero-age main sequence (ZAMS) radius and luminosity are determined as functions of mass and metallicity, utilizing the analytical formulas provided by \citet{tout1996}. 
For the evolution of the star, COMPAS employs the fitting formulas from \citet{1Hurley2000}, which were constructed from the detailed stellar models of \citet{pols1998} for non-rotating stars. The simulation tracks these evolutionary stages using the stellar classification system  derived by \citet{1Hurley2000}. 

The COMPAS version and parameters used in this paper are similar to those described in \citet{Vigna-Gomez2018} and \citet{Howitt2020}. In the following sections, we review for completeness the key binary evolution assumptions relevant to the assembly history of unstable mass transfer systems. For  extensive details on the methodology and implementation of COMPAS, we  refer the reader to \cite{TeamCompas2022,TeamCompas2025}. 

\subsubsection{Initial distributions}
A binary system in COMPAS is defined by its orbital properties: the masses of the initially more massive \textit{primary} ($M_1$) and less massive \textit{secondary} ($M_2$) stars, the semimajor axis ($a_0$), and eccentricity ($e_0$). Alternatively, the initial orbital period ($P_0$) can be specified instead of $a_0$. COMPAS provides a set of predefined initial condition distributions, allowing Monte Carlo sampling to draw values for each star or binary system from a specified distribution. We sample $10^6$ binary systems at ZAMS, all at Solar metallicity \citep[$Z = 0.02$;][]{2024A&A...681A..57B}, ensuring consistency with the metallicity of the default stellar models used in COMPAS to estimate envelope binding energies \citep{XuLi2010}. Each binary system is then evolved until it reaches its final state.

\begin{figure*}
    \centering
    \includegraphics[width=1.\linewidth]{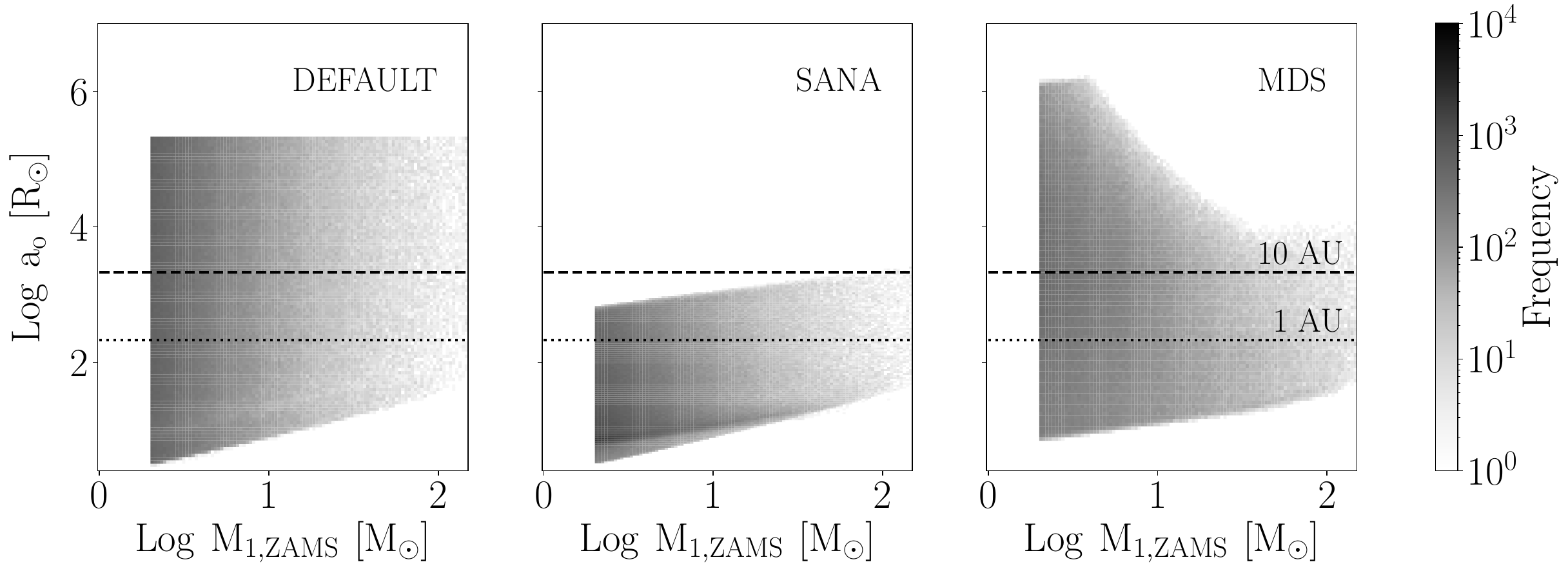}
     \includegraphics[width=1.\linewidth]{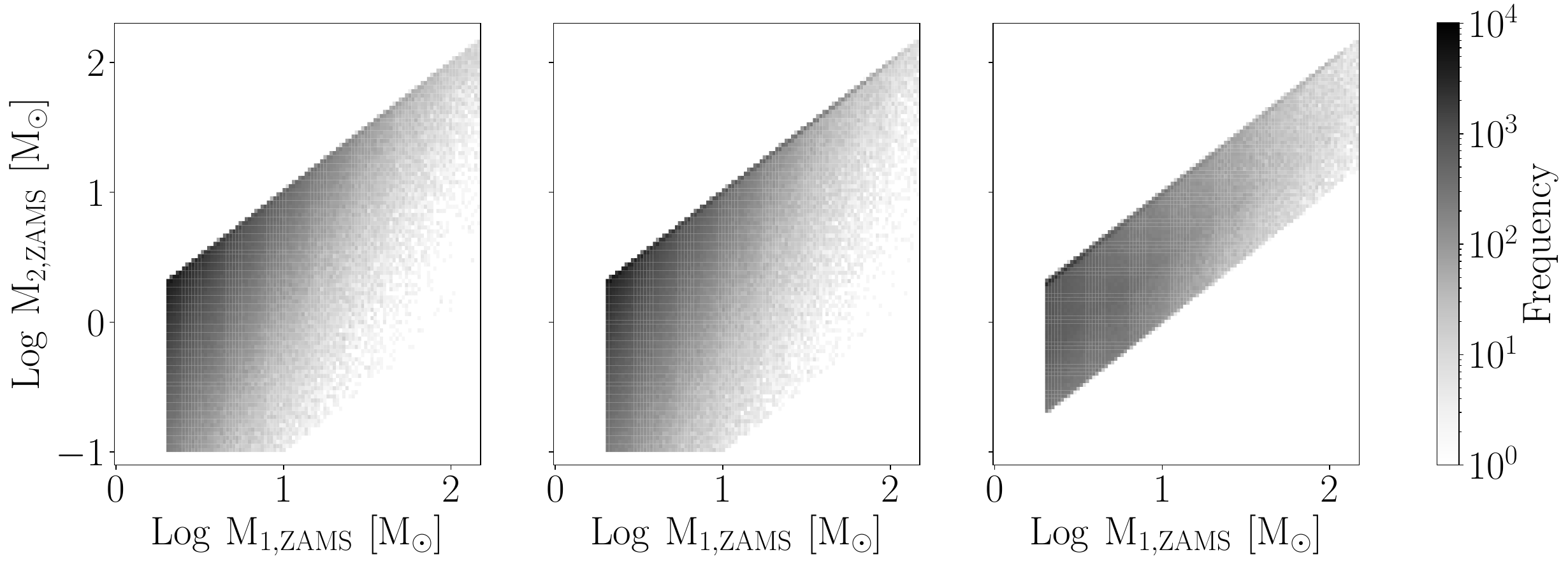}
    \caption{Initial distributions of the binary systems evolved in COMPAS. The three {\it top} panels show the distributions of semimajor axis ($a$ in units of $R_\odot$), and primary  ZAMS  mass ($M_1$ in units of $M_\odot$) for  \texttt{DEFAULT} ({\it left}), \texttt{SANA} ({\it middle}), and  \texttt{MDS} ({\it right}) models. The differences between models are significant and clearly illustrate how model parameter choices critically influence binary population outcomes. The {\it black-dotted} ({\it dashed}) lines denote a binary separation of 1 au (10 au). The {\it bottom} panels shows the ZAMS mass distributions for both $M_1$ (primary) and $M_2$ (secondary) stars.}
    \label{fig:1}
\end{figure*}

In the default settings, the mass of the primary star, $M_1$, is drawn from a Kroupa initial mass function \citep[IMF;][]{kroupa2001} where $M_1 \in [2,150]\, M_\odot$ while the mass of the secondary star, $M_2$, is determined by the initial mass ratio ($q=M_2/M_1$), which is drawn from a flat distribution \citep{Tout1991,Mazeh1992,GoldbergMazeh,KobulnickyFryer2007, 
Sana2012,2014ApJS..213...34K} where $q \in [0.01,1.0]$.
The initial semimajor axis, $a_0$, of a binary star is chosen independently of the masses using a log-uniform distribution where $a_0 \in [0.01,1000]$ au. Unless stated otherwise, COMPAS assumes that orbits are circular at birth, with distributions of the orbital parameters that are independent of each other and metallicity. 
In what follows, we refer to this model as \texttt{DEFAULT}. 

Other distributions for initial masses and orbital configurations are available in COMPAS. For this study we make use of two additional initial parameter distributions for the sampling of our initial binary systems. The first one is derived from \citet{Sana2012} for the uncorrelated distributions of $q \in [0.01,1.0]$ AU, $P_{0} \in [2,10^3]$ days, and $e_0$, which we refer to as \texttt{SANA}. And the second one is derived from \citet{moe2017} for the correlated distributions  $M_1$, $q$, $a_0$ and $e_0$, which we refer to as \texttt{MDS}. For the latest implementation of \texttt{MDS} in COMPAS, we direct the reader to \citet{2025PASA...42...12R}.

Figure~\ref{fig:1} shows the semi-major axis and primary mass for the initial distributions in the \texttt{DEFAULT}, \texttt{SANA}, and \texttt{MDS} models.
When compared to the \texttt{DEFAULT} and \texttt{MDS} models, \texttt{SANA} predicts a steeper distribution of orbital periods and a higher proportion of short-period binary systems. This results in a significantly enhanced fraction of interacting systems. 
In the case of \texttt{MDS}, the period distribution for Sun-like binaries at ZAMS peaks at $\log P_0 ({\rm {days}})\approx 5.0$, corresponding to an orbital semimajor axis of approximately 50 au. 
As a result, a much smaller fraction of systems experience significant alterations from binary evolution. In contrast, the peak for O-type and B-type companion distributions  at ZAMS occurs at smaller semimajor axes (about 10 au), which aligns well with \texttt{SANA}'s predictions. 
In contrast to \texttt{MDS}, the \texttt{SANA} distribution shows no preference for equal-mass binaries. 

\subsubsection{Interacting binaries}\label{sec:bse}
Mass transfer is possibly the most crucial process in interacting binaries, as it modifies both the stellar members and their orbital properties \citep[e.g.,][]{Podsiadlowski1992,2001ASPC..229..239P}.

When two stars in a binary  transfer material, both mass and angular momentum are redistributed. The orbital angular momentum of a gravitationally-bound binary with non-rotating components can be written as\aat{:}
\begin{equation}
    J_{\rm orb} = \mu \sqrt{GM_{\rm tot}a(1-e)},
\end{equation}
where $\mu = M_1M_2/M_{\rm tot}$ is the reduced mass, $a$ is the separation, $e$ is the eccentricity and  $M_{\rm tot} = M_1+M_2$.  
When mass is transferred from one star to another, or is completely lost from the system, the angular momentum of the binary is redistributed and may change.
By its very nature, mass transfer, irrespective of whether it is dynamically stable or unstable, determines the different evolutionary paths and ultimate fates of the system.

In stellar binary systems, interactions between the stars often result in mass exchange. When one star (the donor) expands beyond its Roche lobe, its companion (the accretor) begins to accrete material from the donor’s outer envelope. The donor is not always the primary star; in some cases, the secondary may gain enough mass and later become the donor in a subsequent mass-transfer phase. Hereafter, we denote the mass and radius of the donor as $\rm M_d$ and $\rm R_d$, and those of the accretor as $\rm M_a$ and $\rm R_a$, respectively.

The processes governing mass exchange in binaries are complex, and their parameterizations can be intricate. COMPAS adopts a simplified treatment, following the approaches of \citet{Belcynski2002}, \citet{2002MNRAS.329..897H}, and \citet{2014LRR....17....3P}. It tracks the orbital evolution of a binary system during mass-transfer episodes by solving:
\begin{equation}\label{eq:adot}
     \dfrac{\dot{a}}{a} = 2\dfrac{\dot{J}_{\rm{orb}}}{J_{\rm{orb}}} - 2\dfrac{\dot{M}_{\rm{d}}}{M_{\rm{d}}} - 2\dfrac{\dot{M}_{\rm{a}}}{M_{\rm{a}}} + \dfrac{\dot{M}_{\rm{d}}+\dot{M}_{\rm{a}}}{M_{\rm{d}}+M_{\rm{a}}} + \dfrac{2e\dot{e}}{1-e^2} .
\end{equation}

The default assumption in COMPAS is that the binary system will be circularized instantly at the onset of Roche Lobe overflow \citep[RLOF, e.g.,][]{Counsleman1973,Zahn1977,Zahn2008,VerbuntPhinney1995}. 
By defining the fraction of donor mass loss captured by the accretor,
$\beta=-\dot{M}_{\rm{a}}/\dot{M}_{\rm{d}}$ ($0 \leq \beta \leq 1$), and  assuming that all other material  leaves the system with a fraction $\gamma$ of the specific orbital angular momentum, Equation~\ref{eq:adot}  can be rewritten as:
\begin{equation*}
    \dfrac{\dot{a}}{a} = -2\dfrac{\dot{M}_{\rm{d}}}{M_{\rm{d}}} \Bigg[1-\beta\dfrac{M_{\rm{d}}}{M_{\rm{a}}}-(1-\beta)\left(\gamma + \frac12\right)\dfrac{M_{\rm{d}}}{M_{\rm{d}}+M_{\rm{a}}}  \Bigg].
\end{equation*} 
COMPAS solves this equation to determine the orbital configuration of a binary system following a mass-transfer episode. 

For the donor’s Roche-lobe radius, $R_{\rm RL}$, COMPAS adopts the approximation of \citet{Eggleton1983}:
\begin{equation}
 \frac{R_{\rm RL}}{a}=0.49 \,\frac{q_{\rm RL}^{2/3}}{0.6\, q_{\rm RL}^{2/3} + \ln\!\left(1 + q_{\rm RL}^{1/3}\right)}
\end{equation}

where, ${q}_{\rm RL} = M_{\rm d}/ M_{\rm a}$. 
As the donor star expands beyond its Roche lobe, mass is transferred to the companion through the first Lagrangian point. This point represents the equilibrium location where the Roche lobes of the two stars meet, defining the regions gravitationally bound to each component.

To evaluate the stability of mass transfer, one typically compares  the radial response of the Roche lobe, $\zeta_{\rm RL}=d \ln (R_{\rm RL})/d \ln (m)$,  against the response of the radius of the donor, $\zeta_{\rm d} = \ln (R_{\rm d})/d \ln (m)$, during the process of mass transfer \citep{PacynskiSienkiewicz1972,HjellmingWebbink1987,apjac6269bib101}. In cases where mass transfer leads to the star further surpassing its Roche lobe ($\zeta_{\rm d} < \zeta_{\rm RL}$, the resultant mass transfer becomes unstable (or if both stars fill their Roche lobes near-simultaneously).COMPAS approximates the logarithmic radial response of the donor, $d \log(R_{\rm d})/d \log(m)$, depending on the donor’s stellar type \citep{1Hurley2000}. Following the default COMPAS assumptions, $\zeta_{\rm d} = 2$ for core hydrogen-burning stars and $\zeta_{\rm d} = 6.5$ for Hertzsprung-gap stars \citep{Vigna-Gomez2018,TeamCompas2022}. The radial response to adiabatic mass loss for all evolved stars, with hydrogen envelopes, beyond the Hertzsprung gap is modeled using polytropic models for deeply convective stars \citep{HjellmingWebbink1987,apjac6269bib101}. Mass transfer from exposed helium cores is assumed to be always dynamically stable \citep{Vigna-Gomez2018}, broadly consistent with results from detailed stellar models \citep{2010ApJ...717..724G,2024arXiv241117333G}.

For core hydrogen-burning stars, these assumptions yield a critical mass ratio for unstable mass transfer of $q = q_{\rm RL}^{-1} = M_{\rm a}/M_{\rm d} \lesssim 0.58$ for fully conservative and $q \lesssim 0.44$ for fully non-conservative transfer. For Hertzsprung-gap stars, the critical ratio is $q \lesssim 0.26$ for fully conservative and $q \lesssim 0.21$ for fully non-conservative mass transfer. For more information on the impact of angular momentum redistribution during mass transfer episodes refer to \citet{2023ApJ...958..138W}.
 
When mass transfer becomes unstable, the donor’s outer layers can engulf the accretor, forming a CE \citep{apjac6269bib71}. Drag within the envelope causes the secondary’s orbit to shrink, reducing the separation between the donor’s core and the accretor while depositing energy into the envelope. In some cases, this process results in a stellar merger; in others, the energy is sufficient to eject the envelope, leaving the secondary in a close orbit around the stripped stellar core \citep[e.g.,][]{2013A&ARv..21...59I}. 

In COMPAS, the standard energy formalism \citep{Webbink1984} is used to calculate CE outcomes. In this work, we will use COMPAS only to identify which systems enter a CE phase, but we will not rely on its framework to determine the outcome. Instead, we will enhance the energy formalism with detailed stellar models, to self-consistently account for the internal structure of the star. We also extend the standard formalism by incorporating tidal disruption of stellar companions to predict CE outcomes more accurately.

\subsection{Stellar Structure Profiles with MESA}\label{sec:zones}
In CE episodes, the donor can be an evolved star beyond the main sequence, typically on the giant branch. Such stars share structural features, including extended, diffuse envelopes surrounding a small, dense core that no longer burns hydrogen. However, envelope structure can vary significantly with stellar mass and age \citep[e.g.,][]{2020ApJ...899...77E}. To capture this diversity, we employ the Modules for Experiments in Stellar Astrophysics (MESA) code \citep{apjac6269bib79,apjac6269bib80,apjac6269bib81,paxton2018,apjac6269bib82,2023ApJS..265...15J} to construct two comprehensive grids of one-dimensional stellar structure profiles. The primary grid spans initial masses from 0.08 to 100 $M_\odot$ at $Z = 0.02$, providing a broad, population-scale coverage.

For low\nobreakdash- to intermediate\nobreakdash-mass stars 
($0.08-7.9\,M_\odot$), we adopt the \texttt{7M\_prems\_to\_AGB}
MESA test suite, which evolves stars from the pre-main 
sequence through the asymptotic giant branch (AGB) phase. This setup uses a mixing length of 1.73 and incorporates the Reimers and Blocker RGB and AGB wind schemes, respectively.
For intermediate-mass ($9–13\, M_\odot$) and 
high-mass ($16–100\,M_\odot$) stars, 
we use the \texttt{12M\_pre\_ms\_to\_core\_collapse} 
\citep{2016ApJS..227...22F,2024arXiv240807874H} and \texttt{12M\_pre\_ms\_to\_core\_collapse} \citep{2024arXiv240918189A,2023MNRAS.522..438S} test suites, respectively, evolving stars from ZAMS to core collapse.

To characterize the evolutionary stage of donor stars, we track the central mass fractions of key isotopes (\texttt{h1, he4, c12, o16, n14}) alongside the helium (He), carbon-oxygen (CO), and iron (Fe) core masses. 
We define three evolutionary phases, Zones~I, II, and III, based on the composition and formation of CO or Fe cores, summarized as follows:
\begin{itemize}
    \item Zone~I spans evolution from the ZAMS to the terminal-age main sequence (TAMS). It is identified by the formation of a helium core (\texttt{he\_core\_mass} $\geq 0.01\,M_\odot$) and central hydrogen depletion (\texttt{center\_h1} $\leq 0.01$).
    
    \item Zone~II extends beyond the TAMS, with an upper boundary defined by a mass-dependent, multi-tiered criterion:
    \begin{itemize}
        \item For $M_{\rm d} \lesssim 3.9\,M_\odot$, the boundary occurs when helium (\texttt{he4}) becomes the dominant central isotope;
        \item For $3.9 \lesssim M_{\rm d} \lesssim 12\,M_\odot$, the boundary is set by the formation of a CO core (\texttt{co\_core\_mass} $> 0.01\,M_\odot$);
        \item For $M_{\rm d} \gtrsim 40\,M_\odot$, the boundary corresponds to the formation of an Fe core (\texttt{fe\_core\_mass} $> 0.01\,M_\odot$);
        \item For $13 \lesssim M_{\rm d} \lesssim 40\,M_\odot$, we approximate the boundary with a linear fit to account for scatter in the onset of CO-core formation.
      \end{itemize}
  
     \item Zone~III comprises all remaining evolutionary stages not included in Zones~I or II and, in practice, corresponds to systems with radii larger than those defined by the Zone~II boundary.
\end{itemize}

\begin{figure*}[h]
    \centering
\includegraphics[width=0.9\linewidth]{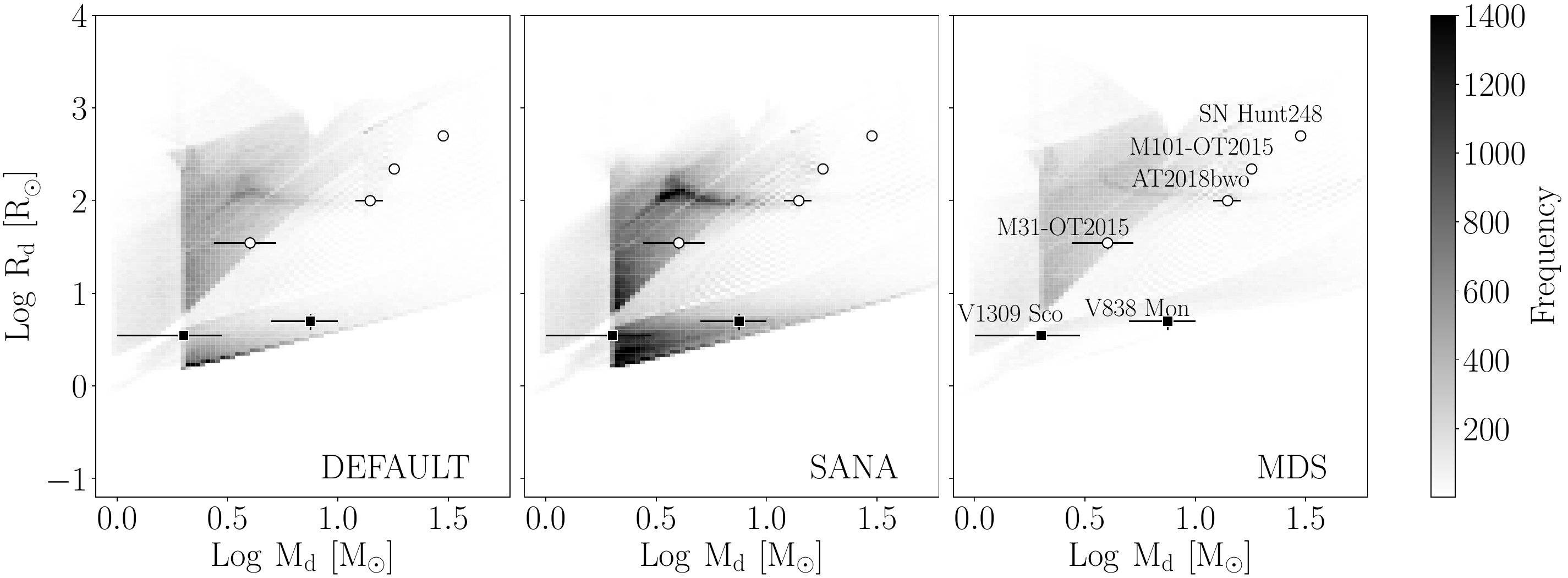}
\includegraphics[width=0.9\linewidth]{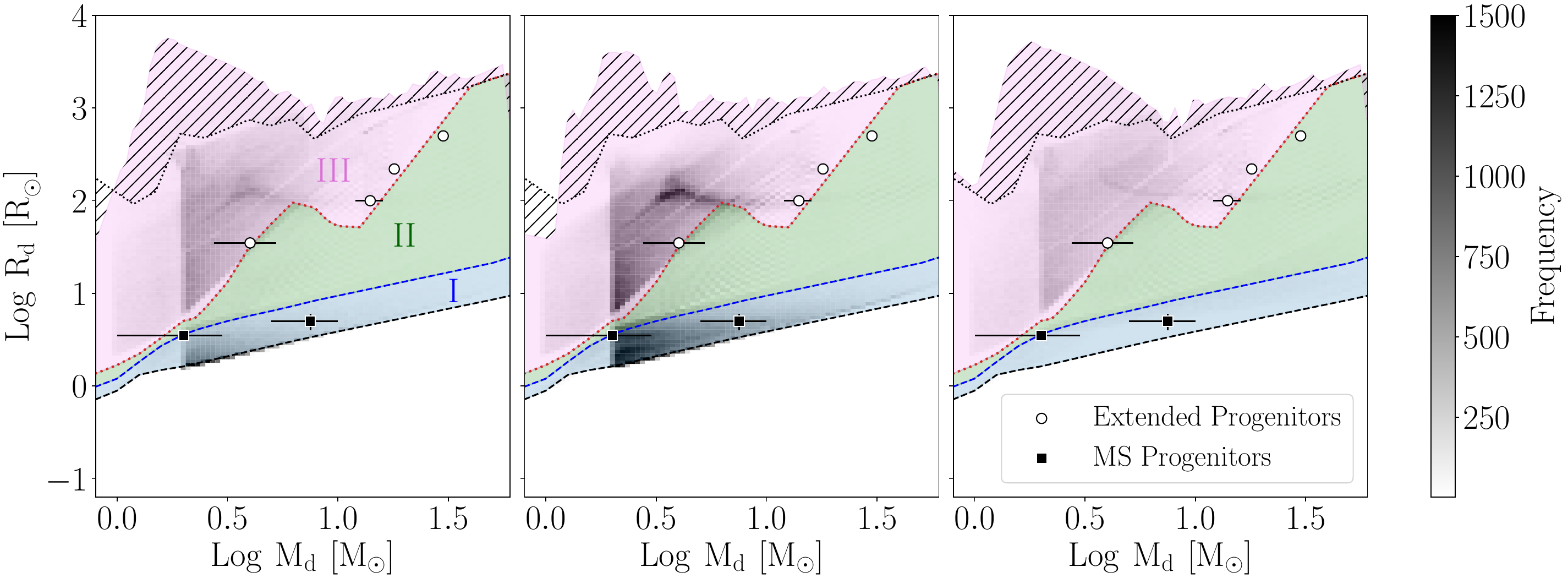}
\includegraphics[width=0.9\linewidth]{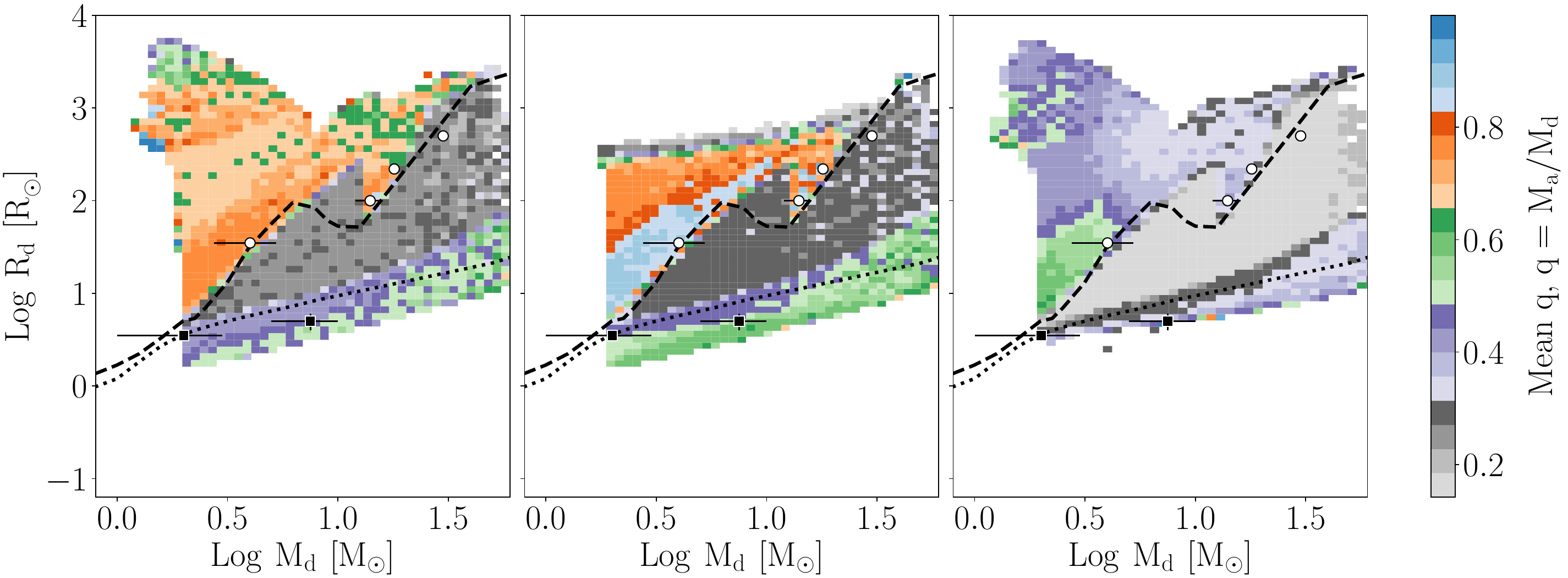}
    \caption{The population of  unstable mass transfer donors in COMPAS. {\it Top} panels show the distribution of donor systems experiencing unstable mass transfer for three different initial conditions. {\it Middle} panels show the  evolutionary zones based on the internal composition and core definitions derived from our MESA models (Section~\ref{sec:zones}): Zone I ({\it blue}), Zone II ({\it green}) and Zone III ({\it pink}). The {\it dashed} and {\it dotted black} lines  indicate the ZAMS and upper radial limit of our MESA grid, respectively. Donors with radii beyond this limit are categorized as hyper-extended (Table~\ref{tab:1}) and are excluded from our analysis.  {\it Bottom} panels  illustrate the mean value of the mass ratio, $q =M_{\text{a}}/M_\text{d}$, of the binary populations. Observed LRNe progenitors with pre-explosion imaging are plotted for comparison \citep{Matsumoto2022}. From left to right (in order of increasing inferred mass), these include: V1309 Scorpii, V838 Monocerotis, M101-OT2015, SN Hunt248, and AT 2018bwo. We further classify these progenitors as either MS progenitors ({\it black squares}; Zone I) or extended progenitors ({\it white dots}; Zones II and III). 
    }
\label{fig:2}
\end{figure*}

\section{The Binary Landscape of Unstable Mass Transfer systems}\label{sec:donors}

\subsection{The Donor Landscape}
To situate our COMPAS findings (Figure~\ref{fig:2}) within the context of observations, we highlight the subset of LRNe  with pre-explosion imaging for which the progenitor’s mass and radius are empirically constrained  \citep[the reader is referred to][for a review]{Matsumoto2022}. This sample includes well-studied transients such as V1309 Scorpii, V838 Monocerotis, M101-OT2015, SN Hunt248, and AT 2018bwo. Together with this sample, Figure~\ref{fig:2} shows the mass–radius distribution of donors immediately prior to the onset of RLOF leading to a CE. To aid interpretation, we map these systems onto the MESA-defined evolutionary phase zones defined in Section~\ref{sec:zones}.  

Donors in the Hertzsprung gap primarily occupy Zone~II (\textit{green}), defined between the {\it black dashed} and {\it dotted} lines in Figure~\ref{fig:2}. These donors are evident when plotting the mean $q$ value of the binary population, whose stability criteria, sensitive to $\zeta_{\rm d}$, change significantly between the TAMS and beyond the Hertzsprung gap (Section~\ref{sec:bse}). Minor differences in the distribution of Hertzsprung-gap donors between COMPAS and MESA arise mainly from variations in stellar evolution treatments. The {\it black dotted} line is the upper radial limit of our MESA grid. Donors with radii beyond this limit are categorized as hyper-extended indicated by the hatched region and are excluded from our analysis. 

Table ~\ref{tab:1} summarizes the population fractions of binary systems across evolutionary Zones I, II and III.  We find that the majority of systems in all three models land in Zone III. A clear lack of donors are found in Zone II, which directly follows from the mass stability criteria implemented in COMPAS. In the sections that follow we show that this Zone II desert is imprinted in the luminosity function of LRNe. Hyper-extended donors are a negligible fraction of the total population.  

\begin{figure}[]
\includegraphics[width=1\linewidth]{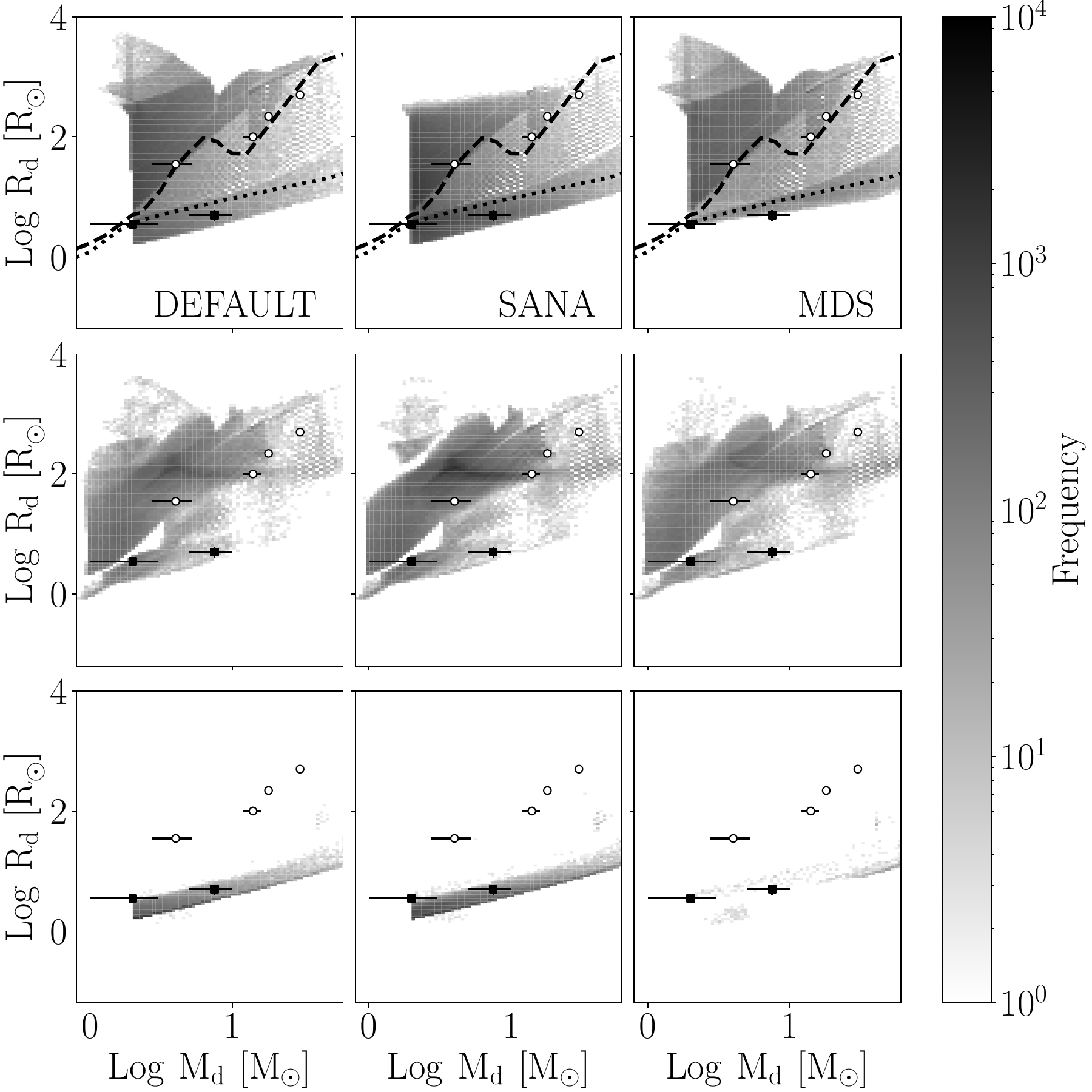}
    \caption{The population of  unstable mass transfer donors in COMPAS for the \texttt{DEFAULT} ({\it left}), \texttt{SANA} ({\it middle}), and  \texttt{MDS} ({\it right}) models.   Primary donors ({\it top} panels), secondary donors (middle panels) and simultaneous RLOF systems ({\it bottom} panels) are shown for comparison. As in Figure~\ref{fig:2}, the constrained mass and radii of observed LRNe with constrained progenitors from pre-outburst imaging are plotted as well as the \textit{dotted} and \textit{dashed} lines demarcating Zones I-II and II-III respectively
    }
    \label{fig:3}
\end{figure}

To better understand binaries at the onset of unstable mass transfer, we examine which component triggers the CE phase and identify systems undergoing simultaneous Roche-lobe overflow (SIMRLOF). Figure~\ref{fig:3} shows the donor landscape in the mass–radius plane, with distributions of primary donors (\textit{top} panels), secondary donors (\textit{middle} panels), and SIMRLOF stars (\textit{bottom} panels). Table~\ref{tab:2} summarizes the population fractions across these three donor categories.

We find that \(\approx 55\)–\(62\%\) of unstable mass-transfer events involve primary donors, while secondary donors account for \(\approx 30\)–\(37\%\), depending on the initial model. These secondary donors often originate from rejuvenated progenitors whose internal structures may not be fully captured by our MESA models \citep[e.g.,][]{Renzo}. The remaining \(\approx 1\)–\(13\%\) of systems undergo near-simultaneous RLOF, with the fraction also sensitive to the initial conditions. The substantial contribution from secondary donors implies that the accretor at the onset of CE can be a stripped core, white dwarf (WD), neutron star (NS), or black hole (BH), which we explore in the following section.

\begin{deluxetable}{lccc}[tbh!]
\centering
\tablewidth{0pt}
\tablehead{
    \colhead{ } & 
    \colhead{FLAT} & 
    \colhead{SANA} & 
    \colhead{MDS} \\
    \colhead{Donor Demographics} & 
    \colhead{(\%)} & 
    \colhead{(\%)} & 
    \colhead{(\%)}
}
\startdata
CE Fraction  & 51.7 & 90.1  & 37.6 \\
 &  &  &  \\
Zone I & 15.8 & 27.4 & 5.9 \\
Zone II & 8.2 & 11.7 & 12.9 \\
Zone III & 69.5 & 58.2 & 76.5 \\
 &  &  &  \\
($\text{M}_\text{d}<$8 M$_\odot$) & 88.8 & 86.5 & 79.7 \\
($\text{M}_\text{d}\geq$8 M$_\odot$) & 11.8 & 13.5 & 20.3 \\
 &  &  &  \\
Hyper-extended donors & 3.1 &  0.38 & 4.6  \\
 &  &  &  \\
\enddata
\caption{Demographics of binary populations undergoing unstable mass transfer, showing the percentage of systems that enter a CE phase, their distribution across evolutionary zones, and donor mass ranges. The hyper-extended donors whose radii exceed the limits of our MESA models constitute a negligible fraction of the total population and are ignored in further analyses. }
\label{tab:1}
\end{deluxetable}

\begin{deluxetable}{lccc}[tbh!]
\centering
\tablewidth{0pt}
\tablehead{
    \colhead{} & 
    \colhead{FLAT} & 
    \colhead{SANA} & 
    \colhead{MDS} \\
    \colhead{CE Event Count} & 
    \colhead{(\%)} & 
    \colhead{(\%)} & 
    \colhead{(\%)}
}
\startdata
Single CE Event & 86.6 & 94.2 &  87.1\\
Double CE Event & 13.4 & 5.8 & 12.9 \\
\cutinhead{Donar Mass Hierarchy}
Primary  & 57.4 & 53.9 & 62.0 \\
Secondary & 32.3 & 32.7 & 36.7 \\
SIM RLOF & 7.8 & 8.7 & 0.38 \\
\enddata
\caption{Demographics of binary populations undergoing one or two CE events and the corresponding mass-donating star for the three models. SIM RLOF indicates simultaneous Roche-lobe overflow.}\label{tab:2}
\end{deluxetable}

\subsection{The Accretor Landscape}
\begin{figure}[]
\includegraphics[width=1\linewidth]{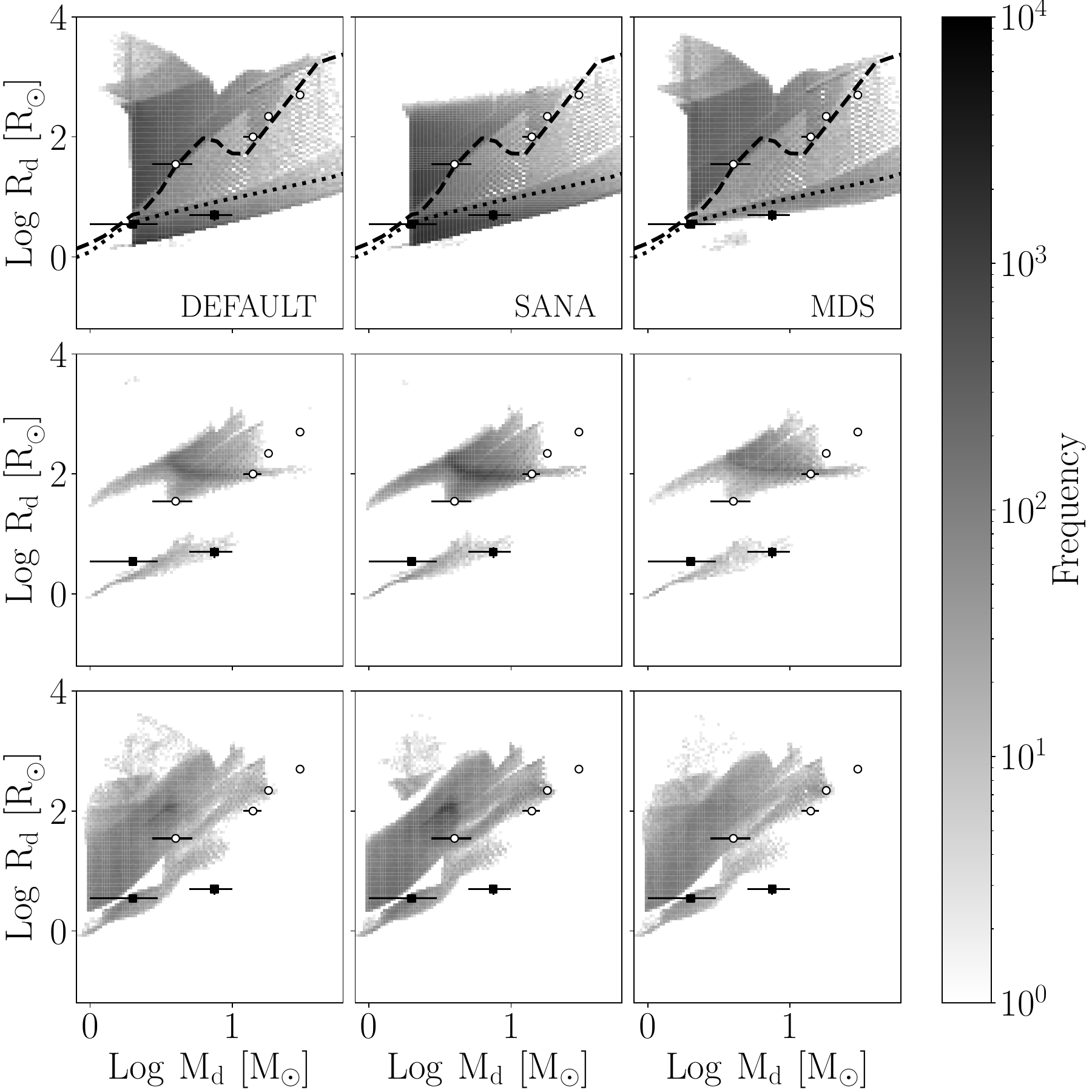}
    \caption{The population of  unstable mass transfer donors  in COMPAS for the \texttt{DEFAULT} ({\it left}), \texttt{SANA} ({\it middle}), and  \texttt{MDS} ({\it right}) models.   Donors are divided based on the accretor stellar type following the  classification of \citet{1Hurley2000}. We show the distribution of donors for  MS  ({\it top}), helium stars ({\it middle}) and WD ({\it bottom}) accretors. As in Figure~\ref{fig:2}, the constrained mass and radii of observed LRNe progenitors with pre-outburst imaging are plotted as well as the \textit{dotted} and \textit{dashed} lines demarcating Zones I-II and II-III respectively.}
    \label{fig:4}
\end{figure}

\begin{figure}[]
\includegraphics[width=1\linewidth]{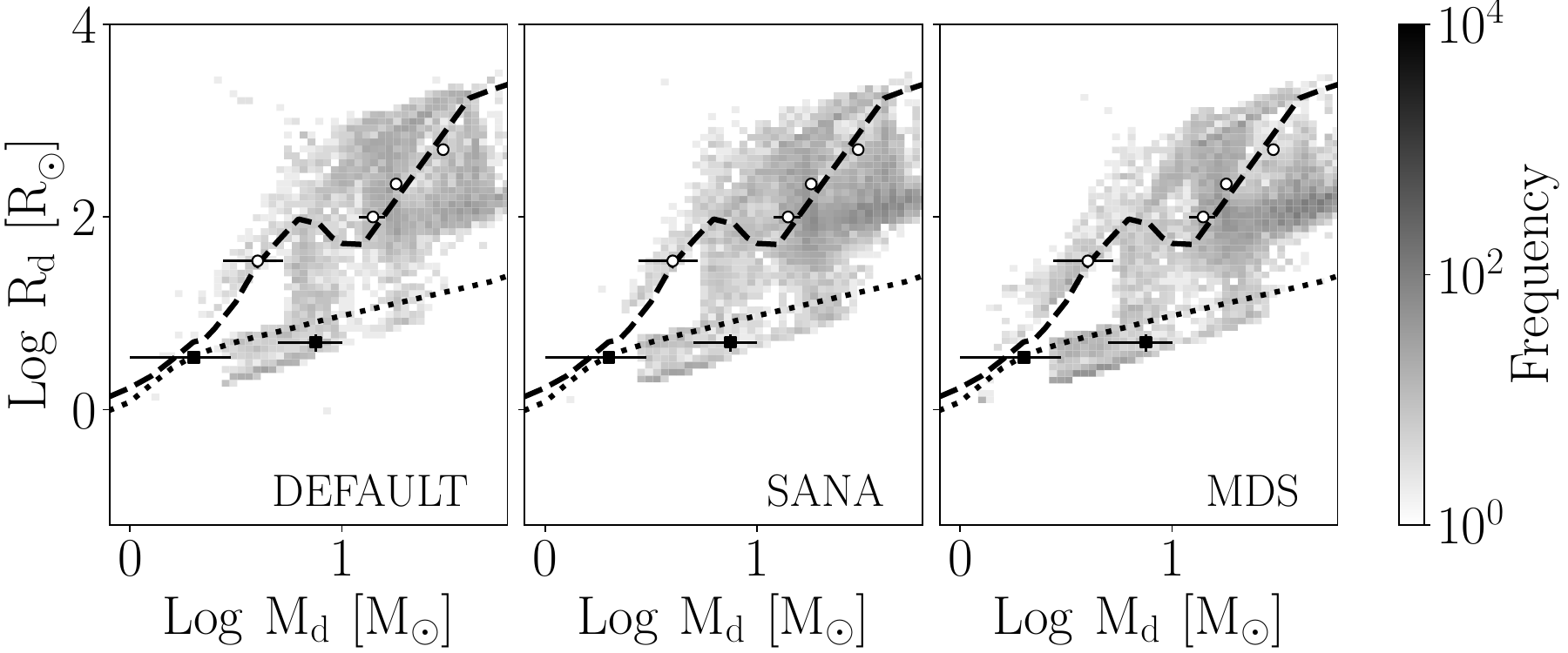}
    \caption{Same as Figure~\ref{fig:4} but for  neutron star and black hole accretors, which are plotted together.
    }
    \label{fig:5}
\end{figure}

As discussed  in Section~\ref{sec:compas}, the stellar type and evolutionary stage of stars in COMPAS are determined based on the nomenclatures from \citet{1Hurley2000}. Table~\ref{tab:3} summarizes the population fractions of accretor types across the three different population synthesis models. Figures~\ref{fig:4} and \ref{fig:5}  show  the accretor landscape 
for stars, WDs,  NS, and BHs, respectively. Once embedded, the accretor will experience gravitational drag, driving the system towards shorter orbital periods until it either merges with the core or ejects the entire envelope. The further evolution of the merger products can result in several exotic and high-energy phenomena, including: collapse to a NS \citep[e.g.,][]{2013A&A...558A..39T} or a BH, formation of an R CrB star \citep[helium-rich giant;][] {2019ApJ...885...27S}, or explosion as a type Ia supernovae \citep[e.g.,][]{2010ApJ...709L..64G,2012A&A...546A..70T}. In the context of NS and BH binaries, outcomes can include  long gamma-ray bursts and exotic merger products like T\.{Z}Os \citep[e.g.,][]{2024ApJ...971..132E,2024ApJ...977..196H, 2025A&A...694A..83N}. 
\begin{deluxetable}{lccc}[tbh!]
    \centering
    \tablewidth{0pt}
    \tablehead{
        \colhead{} & 
        \colhead{FLAT} & 
        \colhead{SANA} & 
        \colhead{MDS} \\
        \colhead{Accretor Type} & 
        \colhead{(\%)} & 
        \colhead{(\%)} & 
        \colhead{(\%)} 
    }
    \startdata
    MS Stars (0-1) & 67.2 & 66.9 &  62.4\\
    WDs (10-12) & 23.8 & 20.5 & 26.0 \\
     &  &  &  \\
    Helium Stars (7-9)  &7.2  &  11.1 & 8.0 \\
    NS+BH (13-14) & 1.3 & 1.2 & 2.8 \\
     &  &  &  \\
    Post-MS (2-6) & 0.6 & 0.4 & 0.8 \\
     &  &  &  \\
    \enddata
    \caption{Accretor stellar type following the nomenclature from \cite{1Hurley2000}}\label{tab:3}
\end{deluxetable}

The most common accretors, in all three models, are main-sequence stars (\textit{top} panel in Figure~\ref{fig:4}), comprising approximately $62\text{--}67\%$ of the total population.
This is not surprising given the large fraction ($54\text{--}62\%$) of donors that are primaries. Helium stars (\textit{middle} panel) represent about $7\text{--}11\%$ of the population and occupy a confined region in mass–radius space. 
These are core helium-burning stars that have lost their hydrogen envelopes through a prior episode of mass transfer or a CE event. Such systems are relevant to rejuvenation \citep[e.g.,][]{Renzo} and chemically peculiar stars \citep[e.g.,][]{1974ARA&A..12..257P}, and they may also inform models of stripped-envelope supernova progenitors \citep[e.g.,][]{2020A&A...637A...6L}.

The next most abundant companions are WDs (\textit{bottom} panel), accounting for roughly $20–26\%$ across models. WD accretors occupy a broad region  in the donor landscape but are primarily abundant in Zone III. These companions have both extended progenitors ($\leq 20M\odot$) and some compact progenitors ($\leq 5M\odot$).

The presence of WDs among both extended and compact donors raises questions about nuclear burning and accretion when a WD becomes embedded in a red giant and merges with its core \citep[e.g.,][]{2008CoPhC.179..184R,2012MNRAS.422.2417D,2017A&A...606A.136L}. It also motivates the study of steady-state accretion solutions for central WDs within red giants. Such systems may form objects akin to Thorne–\.Zytkow  objects \citep{1975ApJ...199L..19T,1992ApJ...386..206C,1995MNRAS.274..485P,2023MNRAS.524.1692F,2024ApJ...977..196H,2024ApJ...971..132E}, particularly if accretion onto the WD powers the object. Helium WDs merging with red-giant cores could produce calcium-rich transients with hydrogen envelopes \citep[e.g.,][]{2018ApJ...858...50F}, illustrating the potential to constrain the nature of unseen companions in stellar merger events.

Across all models, NS accretors constitute \(\lesssim 1.3\%\) of the total population. The \texttt{MDS} model yields the highest fraction (\(1.27\%\)), while the \texttt{DEFAULT} and \texttt{SANA} models produce lower fractions of \(0.58\%\) and \(0.52\%\), respectively. BH accretors are similarly rare, ranging from \(0.6\%\) in the \texttt{SANA} model to \(1.57\%\) in the \texttt{MDS} model. As shown in Figure~\ref{fig:5}, the donors in these systems occupy distinct regions of the donor \(M\)–\(R\) landscape, most commonly within Zone~II.

Although rare, NS and BH accretors are especially important as progenitors of gravitational-wave sources and other high-energy phenomena. Under delayed inspiral conditions, merger outcomes in these systems may also provide a channel for luminous fast blue optical transients \citep[LFBOTs;][] {2001ApJ...550..357Z,2020ApJ...892...13S,2022ApJ...932...84M}.

Recent studies examining common-envelope evolution with NS and BH accretors find that this channel is unlikely to produce stably accreting Thorne–Żytkow objects \citep{2024ApJ...971..132E,2024ApJ...977..196H}. Instead, it may contribute to a subset of X-ray sources and generate transients with observable signatures across the electromagnetic spectrum during and after the common-envelope phase.

\section{Common Envelope Outcomes}\label{sec:mesa}
CE evolution is key to understanding LRNe, yet the process remains poorly understood. While a complete theory of the CE phase is still lacking \citep[e.g.,][]{2013A&ARv..21...59I,2023LRCA....9....2R}, its central role in transient astrophysics, including as the engine for LRNe \citep[e.g.,][]{2024ApJ...963L..35C}, motivates ongoing improvements in both semi-analytical and numerical modeling. The most widely used approach is the energy formalism, which models CE evolution as a comparison between the envelope’s binding energy, $E_\mathrm{bind}$, and the orbital energy, $E_\mathrm{orb}$, released as the secondary spirals inward due to drag \citep{Webbink1984}:
 \begin{equation}
     E_\mathrm{bind} = \alpha \Delta E_\mathrm{orb} ,
 \end{equation}
The change in orbital energy is scaled by an efficiency parameter, $\alpha$, which accounts for potential energy sources and sinks, such as recombination and radiative losses. The formalism predicts that if the orbital energy deposited via shocks is sufficient to overcome the envelope’s binding energy, the entire envelope will be ejected, leaving the secondary in a close orbit around the stripped stellar core \citep[e.g.,][and references therein]{apjac6269bib47}.

While this approach is widely used in rapid population synthesis such as COMPAS, the internal structure of stars varies it cannot self-consistently predict the mass, ejection radius, or velocity of material following a CE episode. Depending on the donor’s evolutionary state, a successful common-envelope ejection (SCEE) may be prevented if the secondary is tidally disrupted, ejecting only part of the envelope \citep[e.g.,][]{2020ApJ...901...44W}. Partial ejection can also occur if the orbital energy is sufficient to lift only the outer layers \citep[e.g.,][]{2025ApJ...979L..11E}. In both cases, the system merges, reducing to a single remnant. To determine which binary configurations result in SCEE, failed ejection, or mergers, we supplement COMPAS with detailed 1D stellar structure from MESA. Following the formalism described below, we construct a grid of $\approx 400$ MESA stellar profiles spaced evenly in log-mass and log-radius, and match each COMPAS donor to the nearest MESA profile using a nearest-neighbor algorithm.

\subsection{Revised Energy Formalism and Tidal Disruption Radius Estimation}
To estimate the radius at which orbital energy is deposited, we use the following form of the standard energy formalism \citep{Webbink1984}, evaluated at each radial coordinate $r$:
\begin{equation}
    E_{\rm bind} (r) =\triangle E_{\rm orb} = \frac{G M_{\rm d} M_{\rm a}}{2R_{\rm d}} - \frac{G M_{\rm d, enc} (r) M_{\rm a}}{2r},
\end{equation}

\begin{figure*}[h!]
    \includegraphics[width=1.\textwidth,trim={120 0 140 0}, clip]{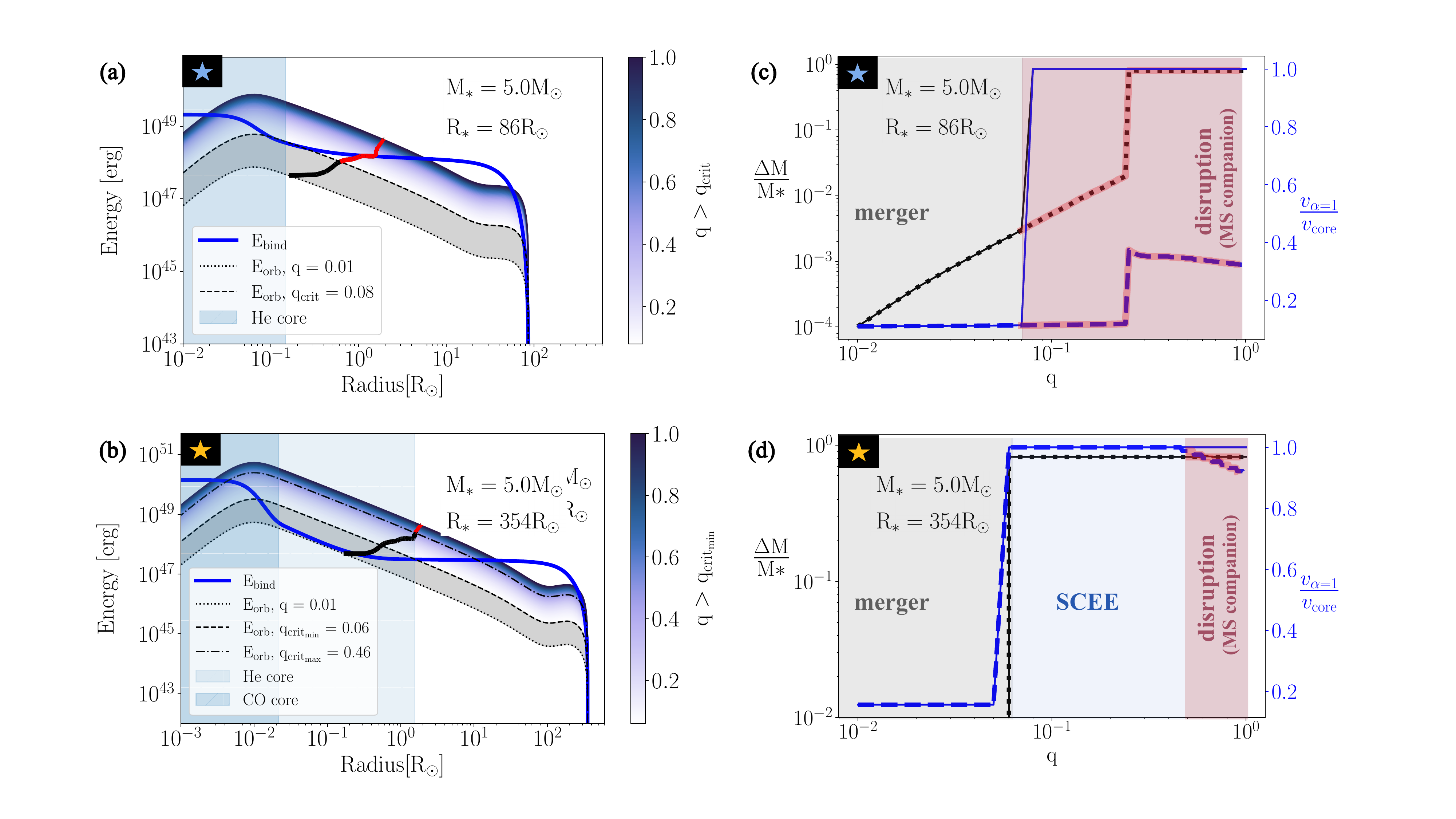}
    \caption{Energy and tidal disruption formalism for CE interactions involving a $5\,M_\odot$ donor at two evolutionary stages ($R_\ast = 86$ and $354\,R_\odot$ represented by the \textit{blue} and \textit{yellow} star symbols respectively). Panels \textbf{a} and \textbf{b} show the energy balance between gravitational binding energy of the envelope and orbital energy deposition for the two donor states and a broad range of companion mass ratios ($0.01\leq q \leq 1$), where $q:=M_{\rm{a}} /M_{\rm{d}}$). The \textit{thick solid black} line that transitions to \textit{red} in both \textbf{a} and \textbf{b} shows the tidal disruption radius, $r_{\rm dis}$, for a MS companion which moves outward as q increases. The \textit{red} transition for this line indicates when this disruption occurs before the stellar companion reaches the required depth to result in a SCEE. The outcomes including ejecta mass and speed estimates for these systems are shown in plots \textbf{c} and \textbf{d} for the $R_\ast = 86$ and $354\,R_\odot$ donors respectively. Here, the \textit{solid} and \textit{dotted} lines represent estimates derived from the standard and revised tidal disruption formalism respectively, with the colors \textit{blue} and \textit{black} indicating the fractional mass and speed estimates respectively. For the $86\, R_\odot$ donor (\textbf{c}) the standard formalism predicts that systems with companions with $q \geq q_{crit}$ would result in complete ejection of the envelope as shown by the sharp increase and plateau of the \textit{solid} lines. In the revised formalism however, all the MS companions that lead to SCEE fail due to disruption. For the $354\, R_\odot$ donor (\textbf{d}), there are companions within a range of q values for which SCEE is possible in the revised formalism. As q increases however,the more massive companions ($q_{\rm crit_{\rm max}} < q \leq 1$) disrupt before the SCEE criteria is met, interrupting the complete ejection of the envelope. These plots show that accounting for the disruption of the companion is necessary since the ejecta estimates and outcomes are significantly different from what the standard formalism would predict.}
    \label{fig:fourpanel}
\end{figure*}

\noindent where $M_{\rm d,enc}(r)$ is the mass of the primary enclosed within radius $r$, and $E_{\rm bind}(r)$ is the gravitational binding energy of the envelope exterior to this radius. We assume all available orbital energy is used to eject this portion of the envelope ($\alpha = 1$). Using this formalism, we identify the radial and mass coordinates where the change in orbital energy exceeds the binding energy of the overlying envelope, allowing us to estimate the ejecta mass. The ejecta velocity is then approximated as the escape velocity at this $\alpha = 1$ coordinate for a given mass ratio.

In the standard picture, if the available orbital energy is sufficient to unbind the envelope down to the core boundary, $E_\mathrm{orb}(r_{\mathrm{core}}) \geq E_\mathrm{bind}(r_\mathrm{core}) $, the system undergoes a successful common-envelope ejection (SCEE), leaving a post-CE binary composed of the stripped core and a surviving companion. If only part of the envelope is removed, the outcome is a partial ejection that leads to a merger. The minimum companion mass ratio required for SCEE is the critical mass ratio, $q_\mathrm{crit}$; companions with $q<q_\mathrm{crit}$ therefore inevitably merge.

In our formalism, if during the dynamical inspiral, the accretor encounters an enclosed density $\overline{\rho}_{\rm enc}$ comparable to half its average density, $\overline{\rho}_{\rm a}$ we assume the companion will experience tidal disruption halting further inspiral and further energy deposition. The inner radius where $ \overline{\rho}_{\rm enc} \approx \frac{1}{2}\overline{\rho}_{\rm a}$ occurs is thus the tidal disruption radius $r_{\rm dis}$ \citep{2020ApJ...901...44W}. If at $r_{\rm dis}$, \( \Delta E_{\text{orb}} \gtrsim E_{\text{bind}}\), the ejecta mass, radius and speed are estimated from that radial coordinate. However, if this condition is not met at $r_{\rm dis}$, but at an outer radial coordinate, $r > r_{\rm dis}$, then ejecta mass, radius, and speed are estimated from that radial coordinate instead. This condition implies that, for failed SCEE systems and certain mergers, neglecting the disruption criterion may result in an overestimation of the ejecta properties.

In Figure~\ref{fig:fourpanel} we show, according to this revised energy formalism,  the various outcomes for a broad range of CE interactions involving an evolved  stellar donor with $M_\ast = 5\, M_\odot$ at two different evolutionary stages $R_\ast = 86\, R_\odot$ ({\it top} panels) and $R = 354\, R_\odot$ ({\it bottom} panels). 
In the \textit{left} panels of Figure~\ref{fig:fourpanel}, \textit{a} and \textit{b}, we show the binding-energy profile of the donor envelope and the orbital energy deposited by a companion, with mass ratios \(q\) ranging from 0.01 to 1, both evaluated as functions of radius. The \textit{gray} region (between the \textit{black dotted} and \textit{dashed} lines) indicates the range in \(\Delta E_{\rm orb}\) for which partial ejection or merger is possible only for \(q < q_{\rm crit}\), where \(q_{\rm crit}\) is the minimum mass ratio required for complete envelope ejection. This threshold is defined by the SCEE condition (\textit{black dashed} line) described in Section~4.1.

The \textit{right} panels, \textit{c} and \textit{d}, present the ejecta mass and velocity as functions of mass ratio for both the standard formalism (\textit{solid} lines) and the new treatment including tidal disruption (\textit{dashed} and \textit{dotted} lines). In panel \textit{c}, for \(q > q_{\rm crit} \approx 0.31\), the standard formalism (\textit{solid} lines) shows a sharp jump in ejecta mass and velocity, signaling the onset of SCEE at the critical mass ratio. As \(q\) increases and companions become more massive, the outcome remains unchanged: the entire envelope is ejected at approximately the escape speed at the core--envelope boundary.

In the new treatment, tidal effects cause more massive main-sequence companions (\(M_{\rm a} = q M_\ast\)) to disrupt at larger radii (\(r_{\rm dis}\)). This prematurely halts inspiral and energy deposition, lowers ejection velocities (\textit{dashed} lines), and significantly reduces the likelihood of successful ejection, leading to smaller ejecta masses (\textit{dotted} lines). 

For the highly extended donor (\(R_\ast = 354\,R_\odot\)) shown in panel \textit{b}, the envelope is more diffuse. Companions with \(q < q_{\rm crit}\) neither disrupt nor deposit sufficient energy to unbind the envelope and therefore sink and merge. The lower binding energy yields a smaller \(q_{\rm crit}\), and in both the standard and revised formalisms SCEE becomes possible for \(q \geq q_{\rm crit}\). However, in our treatment, more massive companions (\(q \gtrsim 0.35\)) become increasingly tidally vulnerable and disrupt just outside the core--envelope boundary, as seen in panel \textit{b}, producing a corresponding drop in ejecta velocity.

Disruption at the core--envelope boundary still results in a merger, but, as shown in panel \textit{d}, the mass ejected in such cases is comparable to that predicted for SCEE under the standard formalism. Thus, even when tidal disruption is included, both stellar mergers and SCEE events can eject similar amounts of mass, with the latter leaving behind the donor core and a surviving companion. This result suggests that only a small fraction of donors retain surviving main-sequence companions after the first CE phase.

\begin{figure*}[ht]
    \centering
    \includegraphics[trim={1cm 1cm 3cm 1cm},clip,width=\textwidth]{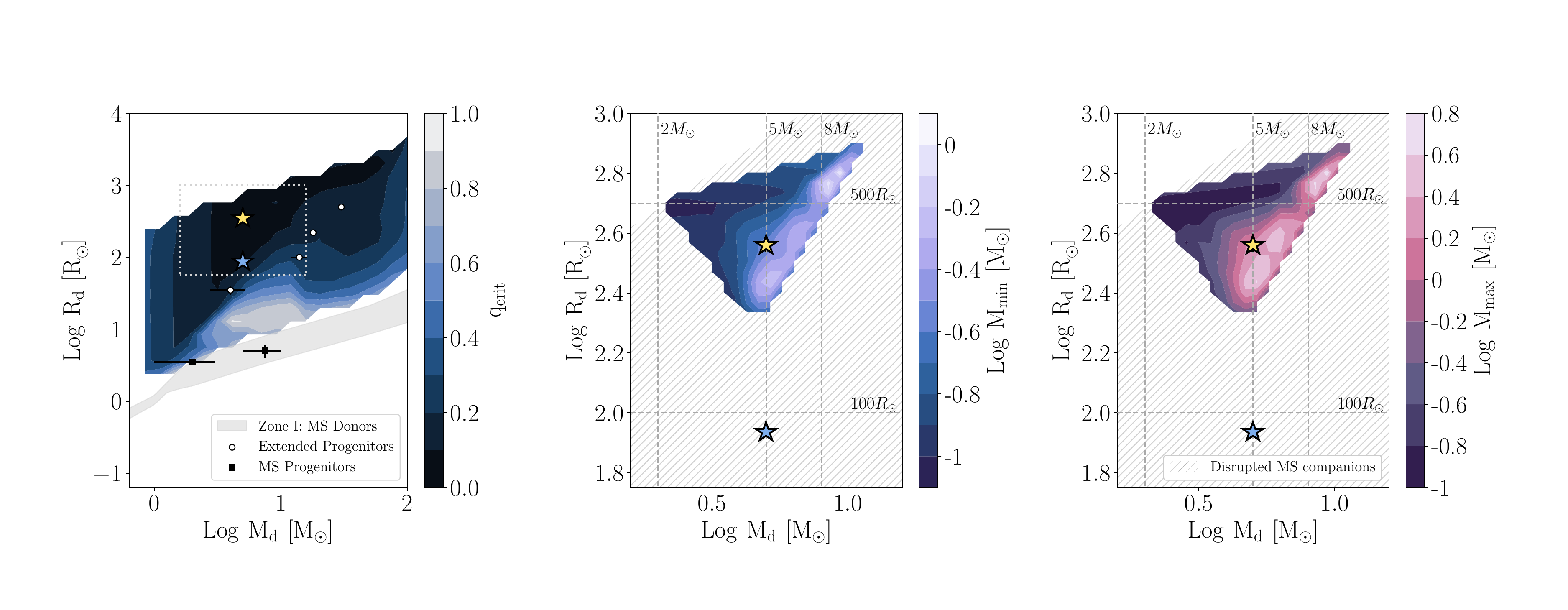}
    \caption{The {\it left} panel shows critical mass ratios ($q_{\rm crit}$) for successful envelope ejection computed using MESA stellar profiles under the standard energy formalism for various companion across $0.01 \leq q \leq1$. The continuous field is interpolated from discrete grid points. Extended progenitors with lower binding energies require lower $q$ values, while compact donors demand significantly higher mass ratios for ejection. The {\it middle} and {\it right} panels show the minimum ($M_{\rm min}$) and maximum ($M_{\rm max}$) MS companion masses capable of surviving the inspiral and ejecting the envelope across the donor $M-R$ landscape. These maps are constructed by interpolating results from our MESA stellar model grid, comparing the donor’s enclosed average density  to half the  average density of a MS companion at the radius where $r = r_{\rm dis}$. Regions that appeared viable for complete envelope ejection under the standard energy formalism are now shown as \textit{gray-hatched} lines indicating that all MS companions would be tidally disrupted before successfully ejecting the entire envelope. Only the colored regions represent donor structures where SCEE  is possible. 
       In all three plots, the {\it blue} and {\it yellow} star symbols represent the two $5\,M_\odot$ donors at $R_\ast = 86$ and $354\,R_\odot$ respectively as introduced in Figure \ref{fig:fourpanel}. }
  
    \label{fig:7}
\end{figure*}

\subsection{Landscape of CE Outcomes}

Figure~\ref{fig:7} illustrates how the likelihood of successful envelope ejection varies across the donor mass-radius landscape for both the standard formalism (\textit{left}) and the revised formalism that considers companion disruption (\textit{middle} and \textit{right}). In the {\it left} panel, we present the interpolated distribution of critical mass ratios ($q_{\rm crit}$) needed to unbind the envelope. In this regime, more extended progenitors, whose envelopes are more weakly bound, require lower $q$ values for complete ejection, whereas more compact donors require substantially higher mass ratios. However, in this formalism, only the mass of the companion is considered, essentially treating it as a point mass. While this assumption suffices for the treatment of a compact companion such as a WD, NS, or BH, it is not the case for a MS companion.
 
This landscape therefore changes significantly when the tidal disruption of the accretor, particularly for MS stars, is taken into account. The \textit{middle} and \textit{right} panels of Figure \ref{fig:7} illustrate the constraints on MS companion masses that can eject the envelope without being tidally disrupted. Specifically, they show the minimum and maximum MS companion masses, respectively, that can both survive the inspiral and supply sufficient orbital energy to unbind the entire envelope. Inspiral from companions less massive than the lower limit cannot unbind the envelope, while those above the upper limit are disrupted before they can transfer sufficient energy, as the donor is disrupted before reaching the radial coordinate where $\alpha=1$. 

As a result, the treatment of all companions under the standard energy formalism (\textit{left} panel in Figure~\ref{fig:7}) is insufficient to define the full parameter space where SCEE is feasible. For many donor profiles, SCEE is only achievable within a bounded range of companion masses. In the {\it middle} and {\it right} panels of Figure~\ref{fig:7}, the same field in the \textit{left} panel is overlaid with \textit{gray-hatched} regions to emphasize that, although the standard formalism predicts successful envelope ejection for these MS companion masses, such companions do not survive the inspiral and instead disrupt. In these regions, no MS companion can simultaneously survive the inspiral and unbind the envelope. These constraints highlight the necessity of accounting for companion structure and have important implications for distinguishing CE outcomes involving MS versus compact companions, particularly for donors near the tip of the giant branch. 

\begin{figure*}[h!]
    \centering
    \includegraphics[width=0.8\linewidth, trim={0 60 0 0}, clip]{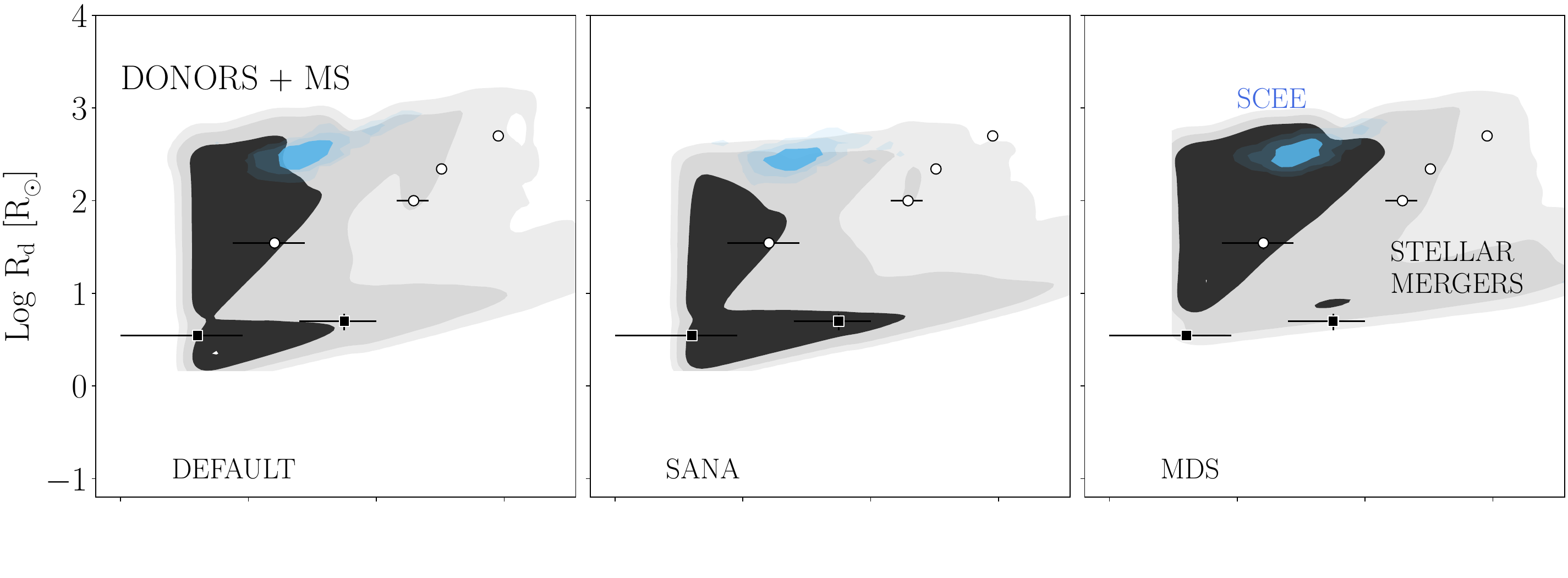}
    \includegraphics[width=0.8\linewidth, trim={0 60 0 0}, clip]{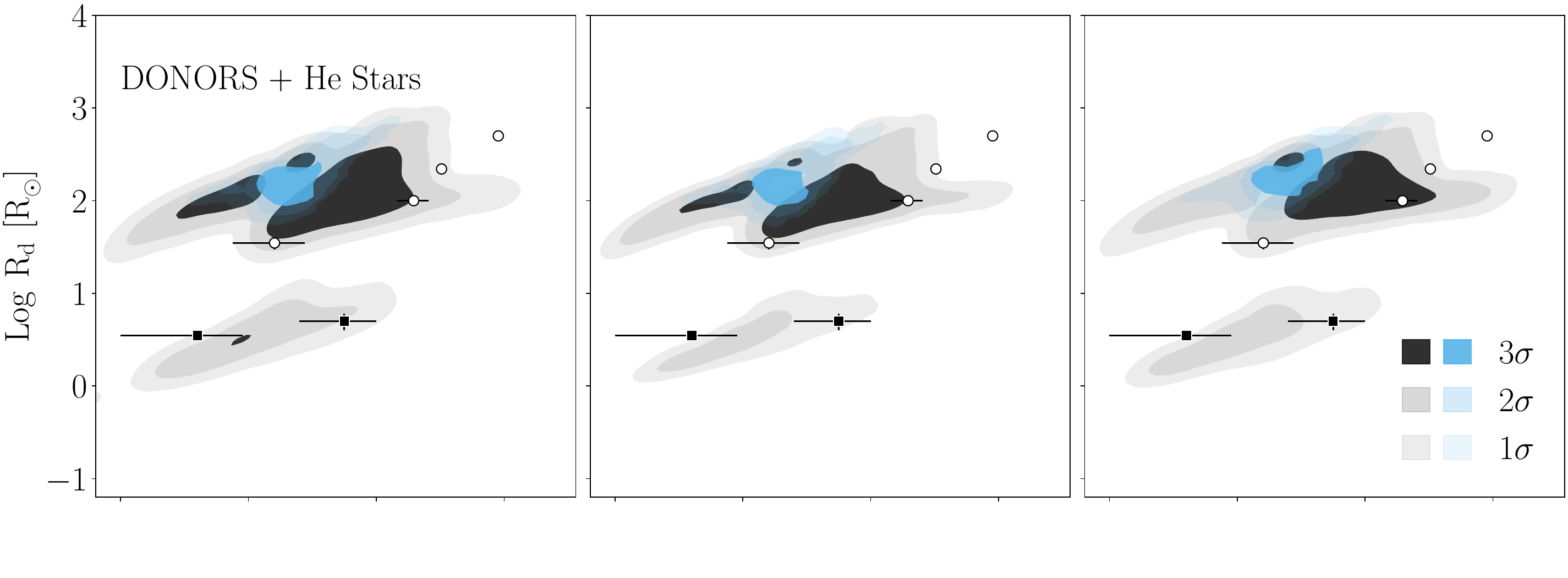}

    \includegraphics[width=0.8\linewidth, trim={0 60 0 0}, clip]{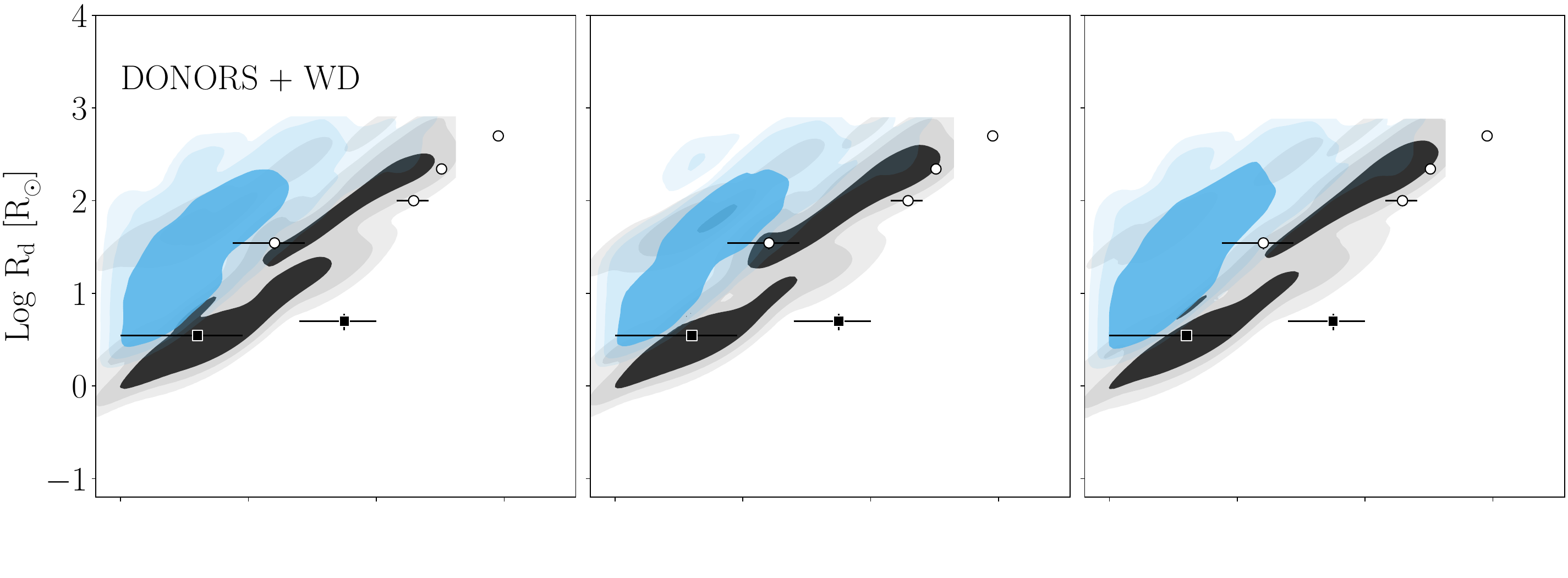}
    \includegraphics[width=0.8\linewidth, trim={0 0 0 0}, clip]{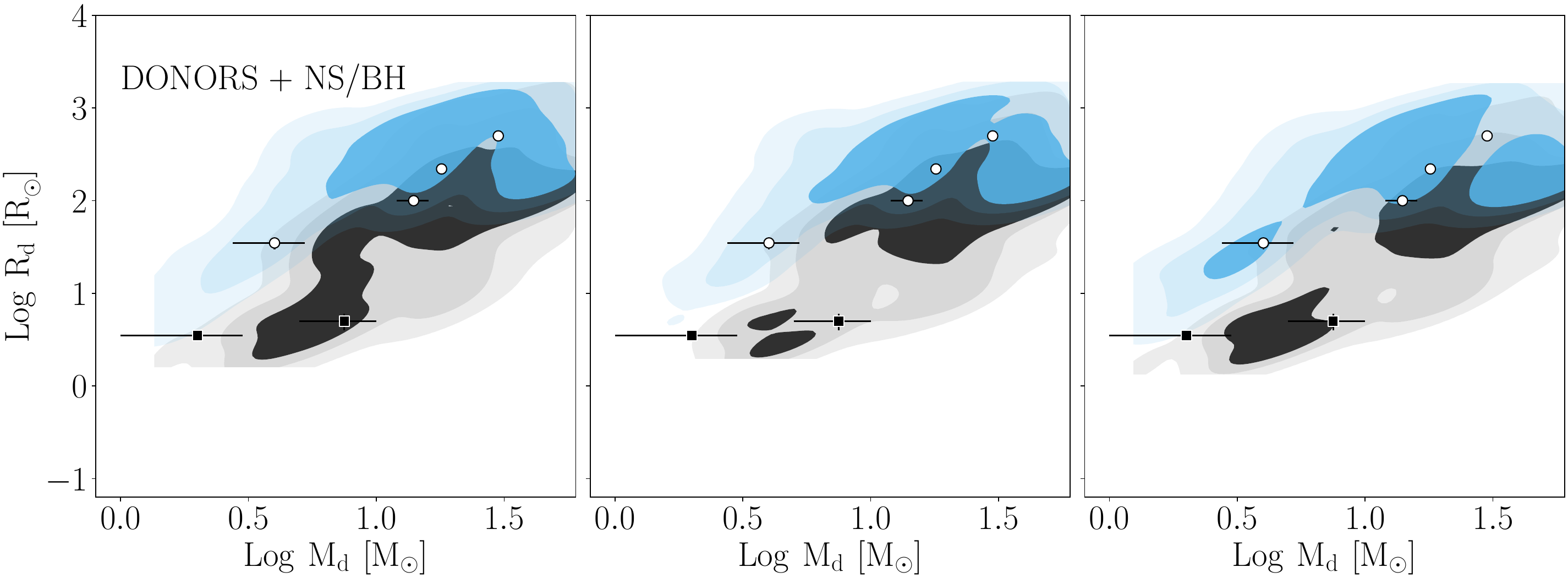}
    \caption{Distribution of CE outcomes in the $M-R$ landscape. The \textit{left, }\textit{middle}, and \textit{right }columns correspond to the \texttt{DEFAULT}, \texttt{SANA}, and \texttt{MDS} models, respectively. From \textit{top} to \textit{bottom}, the panels show COMPAS donors with MS, stripped He stars (Types 7, 8, and 9),  WDs, and relativistic accretors (NS+BH).Contours indicate CE outcomes based on the modified formalism that considers the disruption of the companion: {\it black + gray} 
    contours enclose systems that result in mergers, and {\it blue} contours enclose systems that lead to successful CE ejection. Filled contours show the $1–3\sigma$ density levels, with deeper colors marking areas of greater statistical significance. 
    }
    \label{fig:8}
\end{figure*}

Figure~\ref{fig:8} presents donor \(M\)–\(R\) contour maps of CE outcomes under the revised formalism that incorporates tidal disruption in the three COMPAS populations. From \textit{top} to \textit{bottom}, the panels correspond to donors with main-sequence (MS) companions, stripped He stars, white dwarfs (WDs), and relativistic accretors (i.e., NSs and BHs). The filled contours indicate the \(1\)–\(3\sigma\) density levels, with darker colors denoting regions of higher statistical significance. Stellar mergers are shown in {\it gray}, while the {\it light blue} contours mark systems that undergo SCEE. In the \textit{top} panels (donors with MS companions), the SCEE regions map closely onto the corresponding areas in the \textit{middle} and \textit{right} panels of Figure~\ref{fig:7}.

Taking companion disruption into account drastically reduces the number of MS stars that survive with close-in helium cores. This substantial decrease in survivors fundamentally alters the predicted population and evolutionary pathways of post-common-envelope binaries \citep{2013A&A...557A..87T}. Table~\ref{tab:4} summarizes the outcome percentages for the three  initial binary models, based on the mass and radius limits of the MESA grid and excluding all hyper-extended systems (Table~\ref{tab:1}). As anticipated, inclusion of accretor disruption significantly increases the number of stellar mergers by approximately 30\% across all models.

However, this formalism assumes the donor’s envelope remains static as energy is deposited. In reality, CE episodes are inherently hydrodynamic, with energy and angular momentum transported by waves \citep[e.g.,][]{2025ApJ...979L..11E}. Consequently, energy and angular momentum deposition do not occur under equilibrium conditions \citep[e.g.,][]{2023LRCA....9....2R}. Three-dimensional hydrodynamical simulations by \citet{2024ApJ...977..196H} show that envelope unbinding may proceed slightly more efficiently than predicted by the $\alpha$-formalism. Therefore, our approach is likely conservative and may overestimate the number of mergers. 

\begin{deluxetable}{lcccc}[tbh!]
\centering
\tablewidth{10pt}
\tablehead{
    \colhead{} &
    \colhead{} & 
    \colhead{FLAT} & 
    \colhead{SANA} & 
    \colhead{MDS} \\
    \colhead{Formalism} &
    \colhead{Outcome} & 
    \colhead{(\%)} & 
    \colhead{(\%)} & 
    \colhead{(\%)}
}
\startdata
Standard & Mergers  & 30.0 & 47.1 & 23.8 \\
             & SCEE  & 70.0 & 52.9 & 76.2 \\
 & &  &  &  \\
   Revised & Mergers  & 74.4 & 77.9  & 71.1  \\
              & SCEE & 25.6 & 22.1 & 28.9 \\
& &  &  &  \\
\enddata
\caption{Distribution of CE outcomes (stellar mergers versus SCEE) across models under the standard energy formalism  and the revised formalism that considers the tidal disruption of the stellar companion).}\label{tab:4}
\end{deluxetable}

\section{The Luminosity Landscape of Unstable Mass Transfer}\label{sec:landscape}
\subsection{Mapping Ejecta Estimates to Luminosities}
Since the characteristics of the ejecta associated with LRNe are thought to carry information about the binary progenitor system, it is important to investigate the outcomes of common-envelope events, whether they successfully eject the envelope or lead to a merger.

To connect our analyses with observational signatures of LRNe, we use our ejecta estimates from the MESA grid of donors to predict the resulting optical transients. Light curves of LRNe are often characterized by an initial peak followed by a longer duration plateau that lasts on the order of days to months \citep{Macleod2017,apjac6269bib60,apjac6269bib74}. The plateau phase is thought to be powered from the recombination of hydrogen in the expanding, cooling ejecta as in Type II supernovae \citep{apjac6269bib2}.
Building on extensive work modeling recombination-powered transients \citep{apjac6269bib89, apjac6269bib35}, several studies have applied these principles to LRNe \citep{2013Sci...339..433I, Macleod2017, Matsumoto2022} making it possible to derive analytical approximations of the plateau luminosities from ejecta estimates. 

To estimate the plateau luminosity $L_\mathrm{\rm p}$ and duration $t_\mathrm{\rm p}$ of each event we apply the analytic prescriptions of \cite{Macleod2017} derived using the framework of \cite{2013Sci...339..433I}:
{\fontsize{9}{10}\selectfont
\begin{equation}
\label{eq:lp}
\begin{aligned}
L_\mathrm{\rm p} \approx\ &4.2 \times 10^{37}\, \mathrm{erg\, s^{-1}}
\left( \frac{R_{\mathrm{init}}}{10\, R_\odot} \right)^{2/3}
\left( \frac{\Delta M_{\rm ej}}{0.01\, M_\odot} \right)^{1/3} \\
&\times \left( \frac{\nu_{\rm ej}}{100\, \mathrm{km\, s^{-1}}} \right)
\left( \frac{\kappa}{0.32\, \mathrm{cm^2\, g^{-1}}} \right)^{-1/3}
\left( \frac{T_{\mathrm {rec}}}{4500\, \mathrm{K}} \right)^{4/3},
\end{aligned}
\end{equation}}
and 
{\fontsize{9}{10}\selectfont
\begin{equation}
\label{eq:tp}
\begin{aligned}
t_\mathrm{\rm p} \approx\ &42\, \mathrm{days}
\left( \frac{R_{\mathrm {init}}}{10\, R_\odot} \right)^{1/6}
\left( \frac{\Delta M_{\rm ej}}{0.01\, M_\odot} \right)^{1/3}
\left( \frac{\nu_{\mathrm {ej}}}{100\, \mathrm{km\, s^{-1}}} \right)^{-1/3} \\
&\times 
\left( \frac{\kappa}{0.32\, \mathrm{cm^2\, g^{-1}}} \right)^{1/6}
\left( \frac{T_{\mathrm {rec}}}{4500\, \mathrm{K}} \right)^{-2/3},
\hspace{1cm}
\end{aligned}
\end{equation}}
in which R$_\mathrm{init}$ is the  radius of the donor, $\triangle M_{\rm ej}$ is the ejecta mass, $\nu_\mathrm{ej}$ is the ejecta velocity estimated to be the escape velocity at the radial coordinate where $\alpha =1$, $\kappa$ is the opacity of the ionized gas prior to the passage of the recombination wave, and ${T_\mathrm{rec}}$ is the recombination temperature of the gas.

This framework allows us to translate CE outcomes and ejecta properties into predicted distributions of plateau luminosities and durations, which we directly compare to the observed $L_{90}$ and $T_{90}$ values of LRNe, defined as the luminosity and duration over which 90\% of the radiated energy is emitted \citep{Howitt2020}.

We present estimates of plateau luminosities, $L_{\rm p}$, and durations, $t_{\rm p}$, from Equations~\ref{eq:lp} and \ref{eq:tp}, and analyze the resulting distributions, highlighting how CE outcomes, evolutionary stages, and companion types shape their range and structure.

\subsection{The Influence of CE Outcomes on the Plateau Luminosities and Durations}

\begin{figure}[]
    \includegraphics[width=\linewidth, trim={45 0 60 0}, clip]{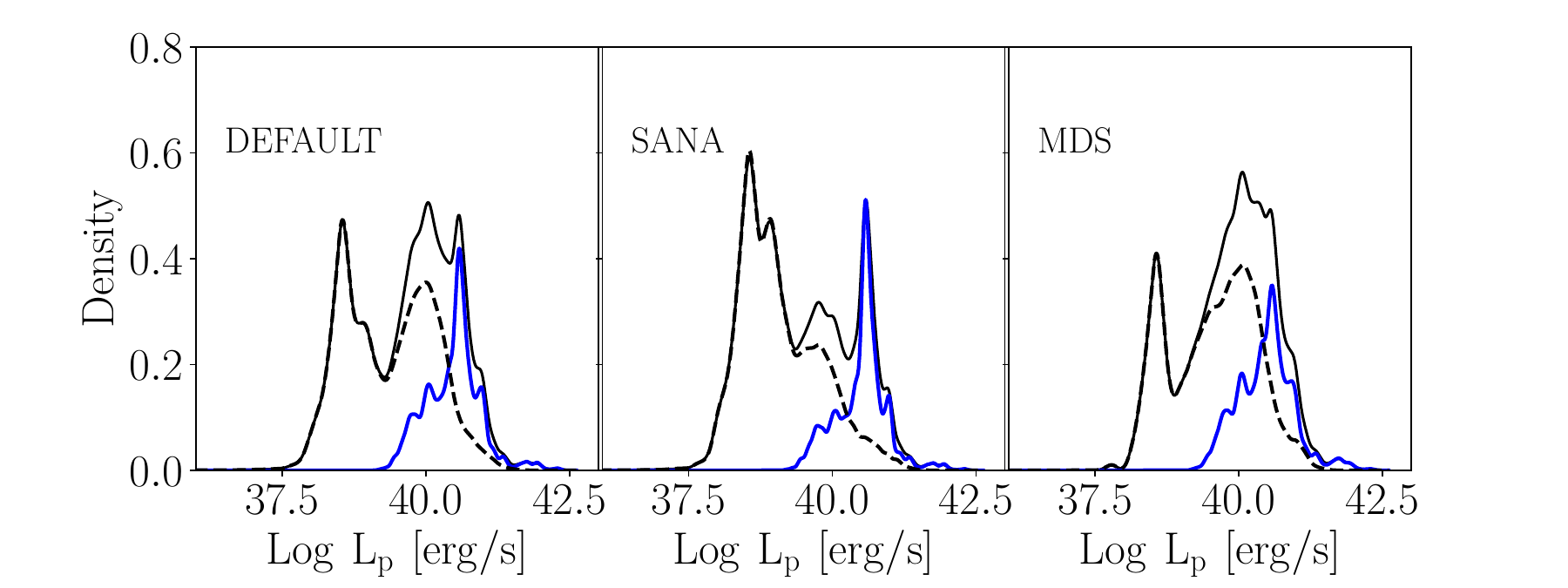}
    \includegraphics[width=\linewidth, trim={45 0 60 0}, clip]{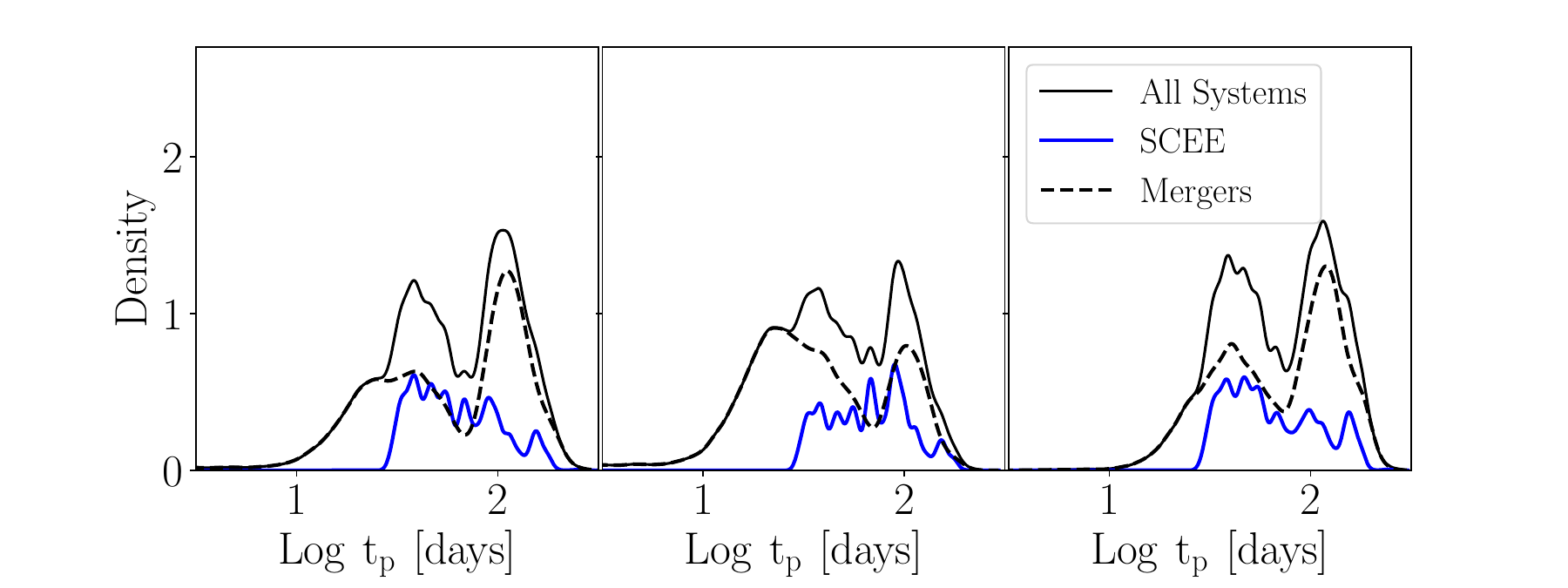}
    \caption{CE outcome contributions to the estimated luminosity and duration plateau distributions ({\it solid-black} lines). Mergers are shown as {\it black-dashed} lines and SCEE as  {\it blue-solid} lines. Our findings reveal a bimodal distribution in plateau luminosities, suggesting that differences in the evolutionary stage of the progenitor system directly influence the resulting plateau luminosity and timescale.
    } 
    \label{fig:outcomes}
\end{figure}

Figure \ref{fig:outcomes} compares the plateau luminosities ({\it top}) and durations ({\it bottom}) of merger ({\it black-dashed} line) and SCEE ({\it blue-solid} line) outcomes to the total distributions ({\it black-solid} line). A broadly bimodal luminosity distribution emerges, with SCEE events producing the bright peak and mergers dominating the faint end. Mergers can contribute to the bright end when companion disruption occurs near the core-envelope boundary. Even without accretor survival, substantial mass ejection can produce luminosities comparable to SCEE events. All three population models show this broadly bimodal pattern, highlighting its robustness across initial conditions. The exact shape and relative prominence of each peak, however, depend sensitively on the population demographics, reflecting variations in binary parameters and evolutionary pathways.

Stellar mergers and SCEE outcomes overlap in both luminosity and timescale distributions. However, stellar mergers dominate the low-luminosity (\(L_{\rm p} \leq 10^{39}\,\mathrm{erg\,s^{-1}}\)) and short-timescale (\(t_{\rm p} \leq 30\,\mathrm{d}\)) regime, suggesting that LRNe observed in this region are most likely produced by stellar mergers.

\begin{figure}[]
    \includegraphics[width=\linewidth, trim={45 0 60 0}, clip]{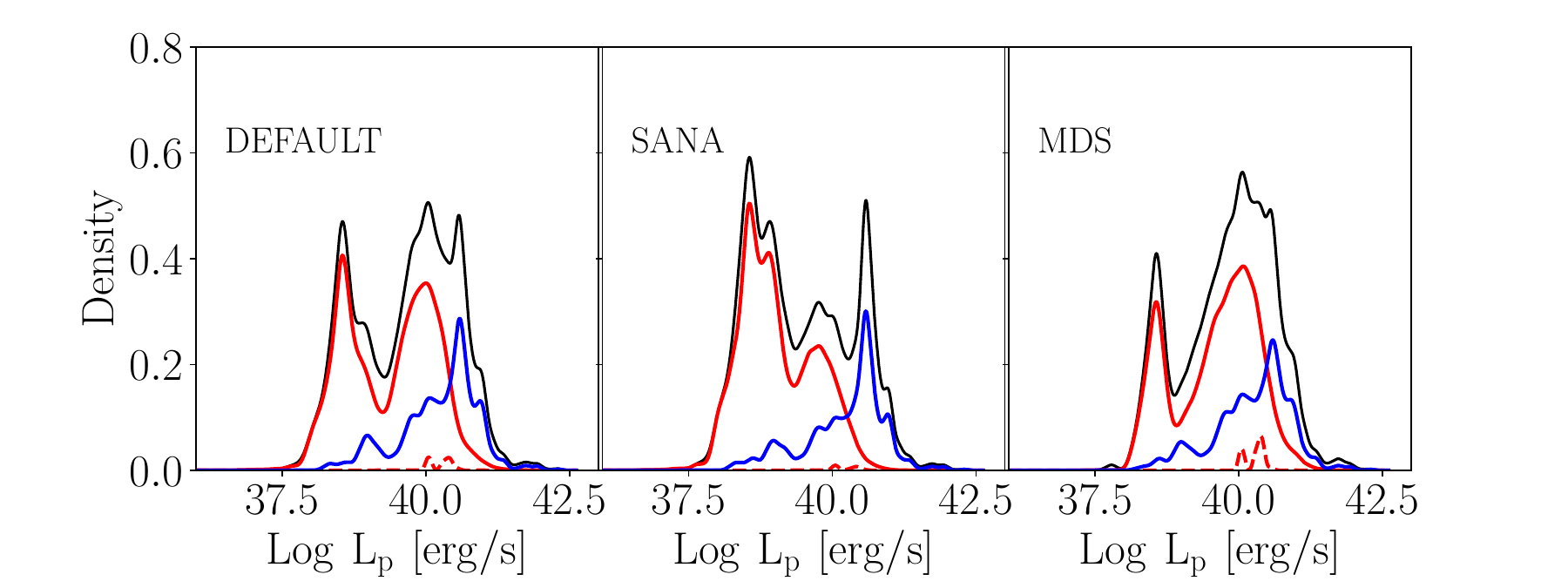}
    \includegraphics[width=\linewidth, trim={45 0 60 0}, clip]{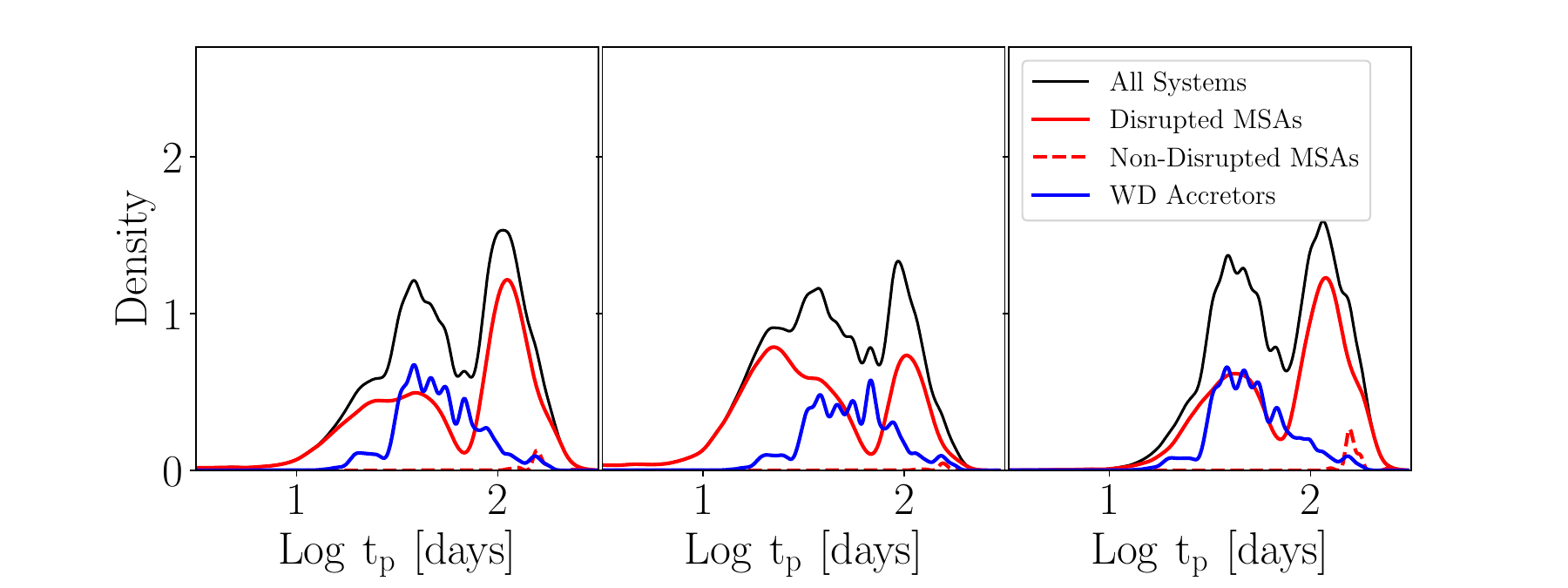}
 
    \caption{The contributions of disrupted MS accretors (MSA) to the plateau luminosity and duration distributions ({\it solid-black} lines). For reference, the contributions from surviving MS ({\it red} lines)  and WD ({\it blue} lines) accretors are also plotted, to aid in direct comparison across different system types. 
    }
    \label{fig:dist}
\end{figure}

The impact of tidal disruption is illustrated in Figure~\ref{fig:dist}, showing the contributions of disrupted ({\it red-solid}) and non-disrupted ({\it red-dashed}) MS accretors to the plateau luminosities ({\it top} panels) and durations ({\it bottom} panels) of the total distributions ({\it solid-black}). For comparison, contributions from WD accretors are shown in {\it blue}. The faint end of the luminosity distribution is dominated by disrupted MS companions, while the bright end is shaped by surviving MS and WD accretors.

\begin{figure}[]
    \includegraphics[width=\linewidth, trim={45 0 60 0}, clip]{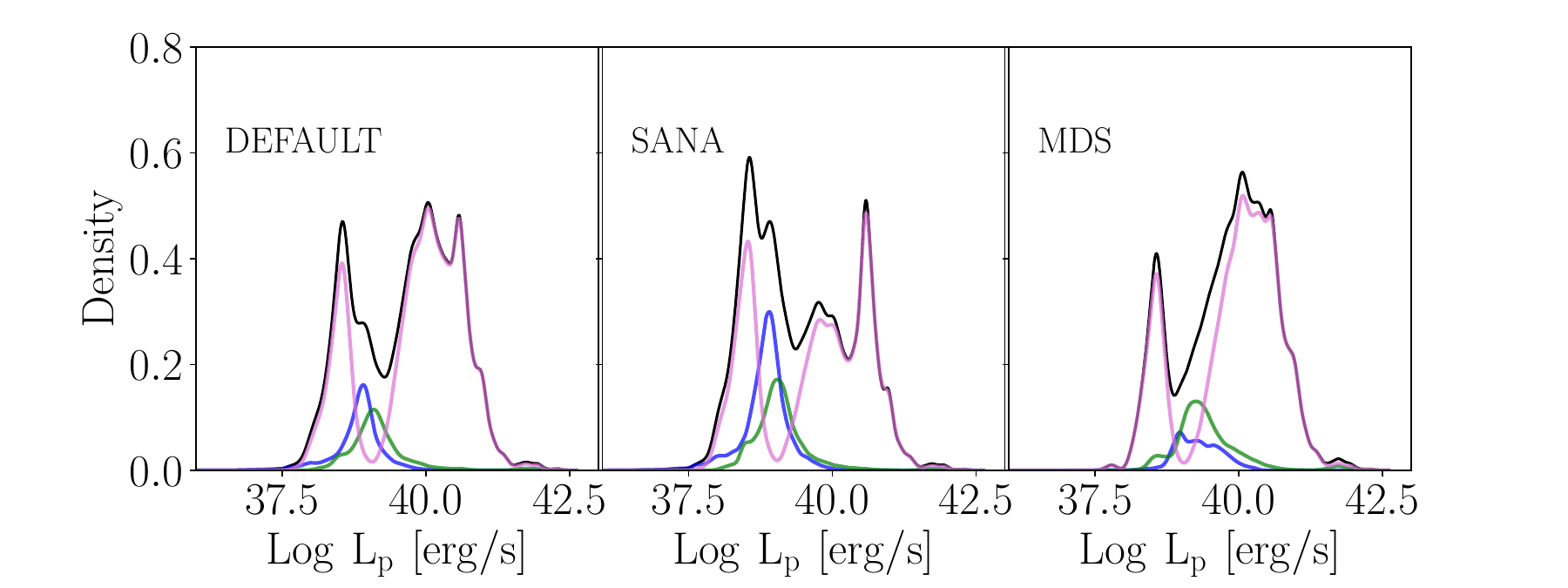}
    \includegraphics[width=\linewidth, trim={45 0 60 0}, clip]{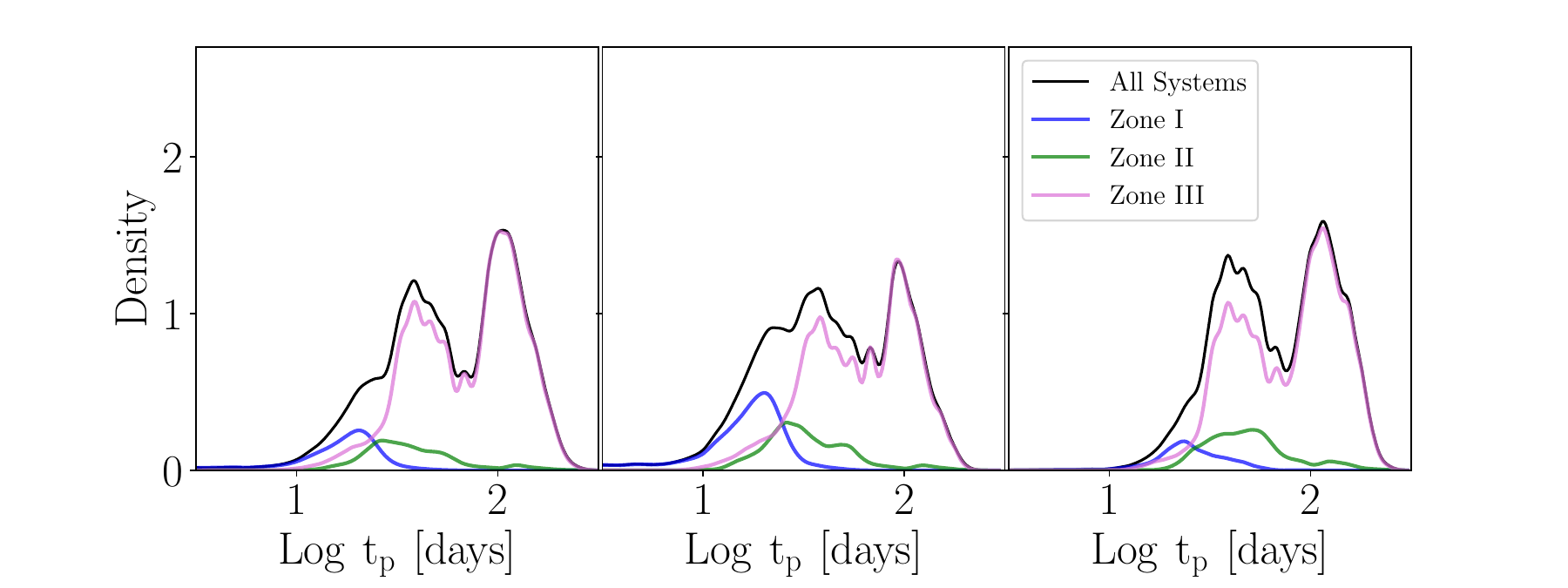}
\caption{Contributions to the plateau luminosity and duration distributions from donors undergoing CE at different evolutionary stages. Following Figure~\ref{fig:2}, donors belonging to Zone I,  II and  III are shown in {\it blue}, {\it green} and {\it purple}, respectively. Our results indicate that evolved progenitors (Zone III) predominantly shape both extreme ends of the luminosity distributions: they account for the bright end through successful complete ejections and the faint end via mergers.}  
    \label{fig:zones}
\end{figure} 

\begin{figure*}[]
    \centering
    \includegraphics[width=01\linewidth]{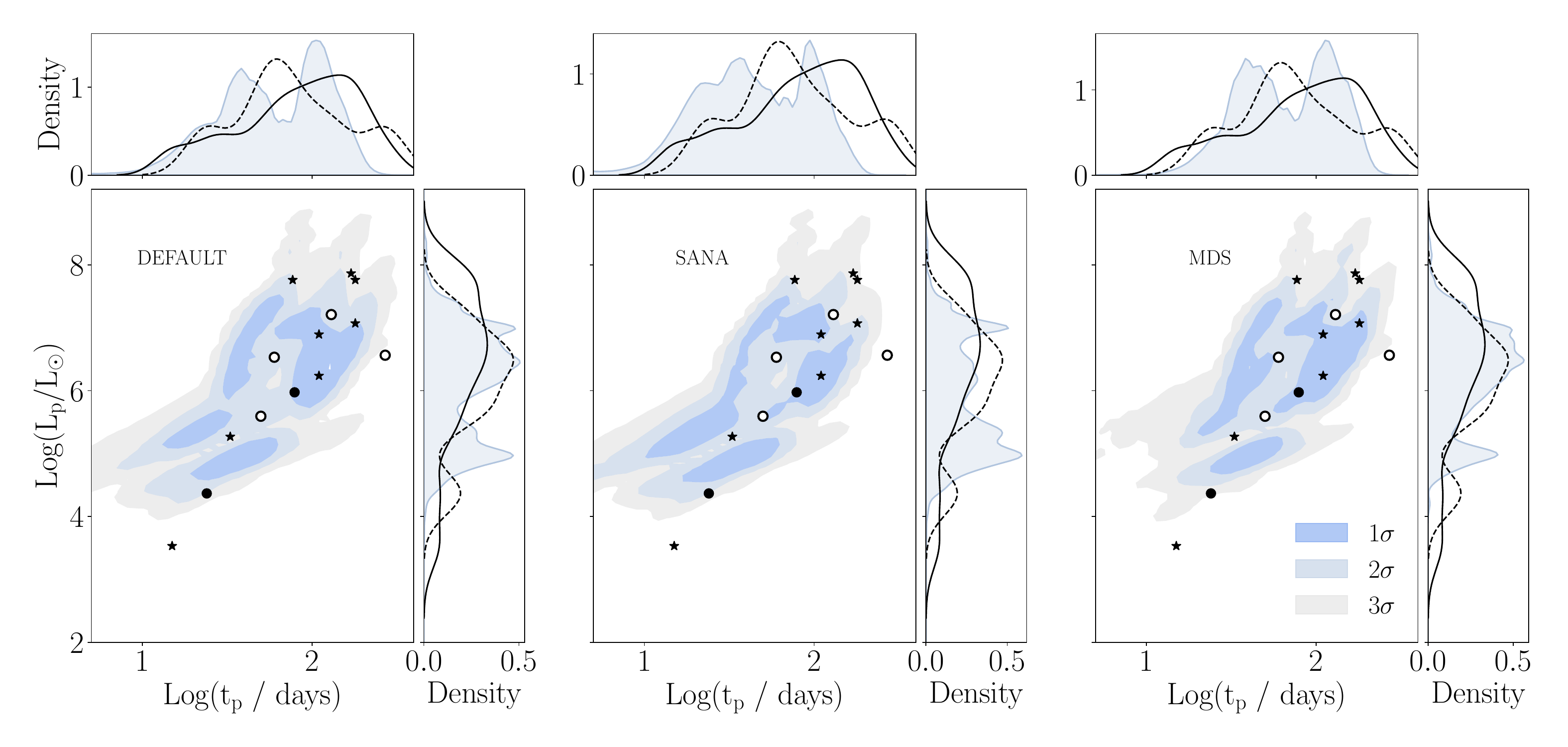}
     
    \caption{Comparison of predicted plateau luminosity ($L_{90}$) and duration ($t_{90}$) distributions from the three COMPAS models, \texttt{DEFAULT} ({\it left}), \texttt{SANA} ({\it middle}), and \texttt{MDS} ({\it right}), with the full observed sample of LRNe compiled by \citet{Matsumoto2022} ({\it dots} and {\it stars}). The subset of observed LRNe with pre-explosion imaging and well-constrained progenitors highlighted by the {\it dots}. The central panels show the joint distributions with filled contours indicating model density levels, while observed LRNe with MS and extended progenitors are marked with {\it black} and {\it white} dots, respectively as shown in Figure~\ref{fig:2}. The {\it top} and {\it right} sub-panels present marginal distributions in timescale and luminosity: filled {\it blue} histograms show model predictions, the {\it solid black} line shows the full observed sample, and the {\it black dashed} line highlights the subset with well-constrained progenitors.
    }
    \label{fig:hst_l90}
\end{figure*}

\subsection{The Imprint of the Donor's Evolutionary State on the Plateau Luminosities and Durations} \label{sec_5.3}

Figure \ref{fig:outcomes} displays the distributions of plateau luminosities (\textit{top}) and durations (\textit{bottom}), sorted by the donor star's evolutionary state and color-coded as in Figure~\ref{fig:2} (Zone I: {\it blue}, Zone II: {\it green}, Zone III: {\it purple}).
Compact and mildly evolved donors (Zones I and II) predominantly produce plateau luminosities in the $10^{38}$–$10^{40}$ erg s$^{-1}$ range. In contrast, extended donors (Zone III) exhibit a pronounced bimodal luminosity distribution, with a fainter peak at $10^{37}$–$10^{39}$ erg s$^{-1}$ and a brighter peak at $10^{39}$–$10^{41.5}$ erg s$^{-1}$. This bimodality persists across all three population models, though the precise shapes and amplitudes vary with underlying demographics (Table~\ref{tab:1}). The distributions indicate that plateau luminosity is a strong tracer of the donor’s evolutionary state. Timescale distributions reinforce this: in the {\tt FLAT} and {\tt SANA} models, Zone~I and II systems typically have plateau durations under 50 days, while Zone~III systems generally exceed 50 days.

\subsection{Comparisons with Observations of LRN transients}
Figure~\ref{fig:hst_l90}, similar to the joint distributions of \citet{Howitt2020}, compares our predicted plateau luminosity and duration distributions from binary population synthesis models with the observed LRN sample compiled by \citet{Matsumoto2022}; with the subsample with pre-explosive imaging and well-constrained progenitors (Section~\ref{sec:donors}) represented as \textit{dots}. The central panels show the joint distribution of $L_{\rm p}$ and $t_{\rm p}$ for the three BPS models: \texttt{DEFAULT} ({\it left}), \texttt{SANA} ({\it middle}), and \texttt{MDS} ({\it right}), with filled contours. Observed LRNe with MS and extended progenitors are marked with {\it black} and {\it white} symbols, respectively. The {\it top} and {\it right} sub-panels present marginal distributions in timescale and luminosity: filled {\it blue} histograms show model predictions, the {\it solid black} line shows the full observed sample, and the {\it black dashed} line highlights the subset with well-constrained progenitors.

Overall, the predicted distributions from the three models are broadly consistent and reproduce the general trends observed in LRNe, although the small number of well-characterized progenitors introduces statistical uncertainty. Observational biases also affect the sample: pre-outburst progenitors are more easily detected if luminous, and LRN discovery depends on survey cadence, depth, and strategy. These factors can skew the observed distributions relative to the intrinsic population, so caution is needed when interpreting model–observation agreement.

While our models reproduce the bulk of LRN observations, they underpredict long-duration plateaus and the most luminous events by 30--60\%, and the most luminous events cannot be explained within a purely recombination-powered framework. This discrepancy is statistically significant, as such systems do not naturally arise from the predicted distributions. The excess of long-timescale transients in the observed sample suggests that mass ejection in these systems is more gradual than in impulsive models, producing lower luminosities and longer plateaus. The most luminous LRNe likely require additional energy sources, such as shock heating, nuclear burning, or accretion, beyond those captured by recombination-based models.

Resolving these discrepancies will require both theoretical and observational advances. Improved modeling of binary interactions, particularly the physics of common-envelope ejection and mass-loss mechanisms, will enhance predictions. Observationally, systematic surveys targeting progenitors and environments of long-duration or luminous LRNe will provide critical constraints. Together, these efforts can refine our understanding of the physical processes driving the diversity of LRN transients.

\subsection{Rethinking Long-Lived LRN outbursts}\label{sec_rethinking}
A key limitation of the current framework is its assumption of impulsive mass ejection \citep{Macleod2017,2025ApJ...979L..11E}. In many systems, envelope ejection may instead occur over multiple dynamical timescales as the companion spirals inward, producing a fainter and longer-lasting plateau. To account for this, we compare the plateau duration, \(t_{\rm p}\), predicted by the impulsive model with the orbital period at the stellar surface, \(t_{\rm k}\), which more realistically characterizes the envelope ejection timescale.

For systems with \(t_{\rm p}/t_{\rm k} > 1\), the impulsive approximation provides a reasonable description of the ejection process and provides an upper limit in the brightness of the LRNe. Conversely, for systems with \(t_{\rm p}/t_{\rm k} \lesssim 1\) the impulsive model overestimates the luminosity and underestimates the true plateau duration.

\begin{figure}[t]
    \centering
        \includegraphics[width=\linewidth, trim={45 0 60 0}, clip]{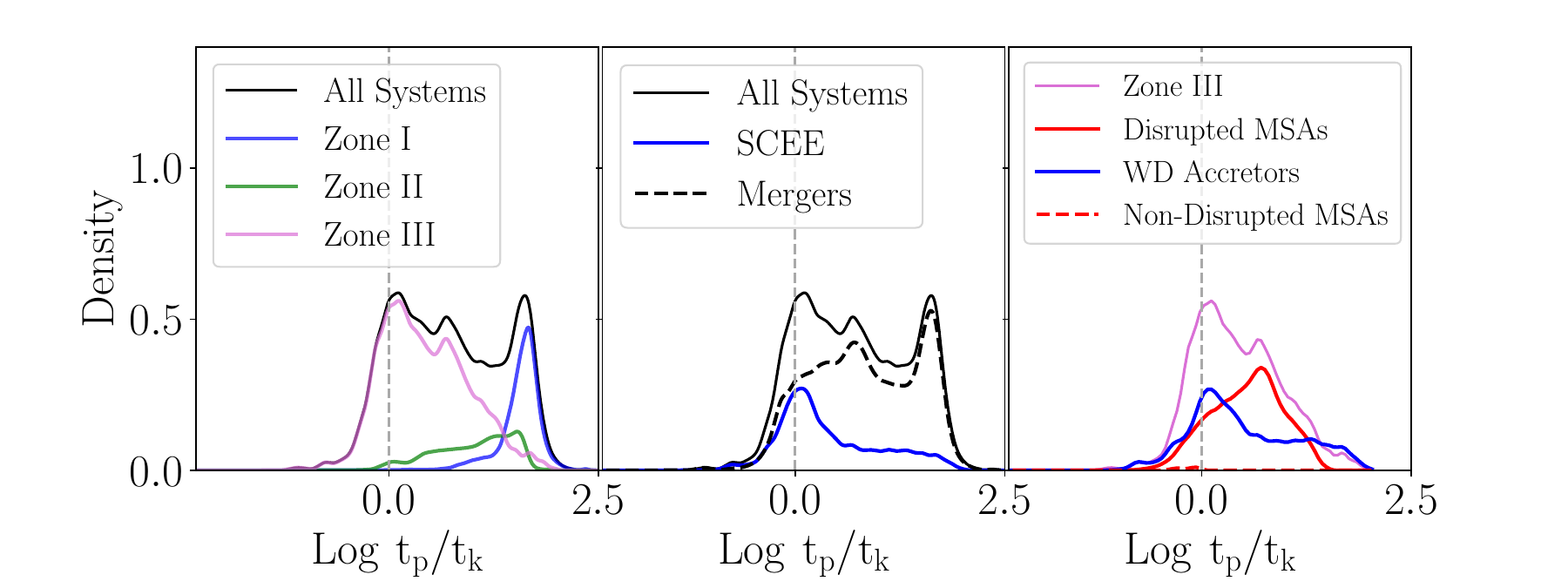}

    \caption{An overview of the timescale ratio $t_{\rm p}/t_{\rm k}$ for the \texttt{SANA} model . The {\it top} row groups events by the evolutionary state of the donor star (Zones I, II, and III. The {\it middle} row distinguishes between CE mergers and SCEEs. The {\it bottom} row delves deeper into Zone III outcomes, highlighting whether the companion is a MS accretor(MSA) or WD, and whether the MSAs disrupt or survive the interaction. 
    }
    \label{fig:13}
\end{figure}

Figure~\ref{fig:13} shows the distribution of \(t_{\rm p}/t_{\rm k}\) for systems in the \texttt{SANA} model across zones (\textit{left}), outcomes (\textit{middle}), and companions of donors in Zone~III (\textit{right}). In the \textit{left} panel, systems with \(t_{\rm p}/t_{\rm k} \lesssim 1\) are predominantly found in Zone~III. 
In these cases, the impulsive approximation offers a poor approximation, and instead the ejection of the envelope would likely proceed over multiple orbital periods.
The resulting luminosities may be dimmer, and plateau durations longer than predicted, potentially reconciling the model under-predictions for the brightest events. By contrast, systems in Zones~I and~II primarily have \(t_{\rm p}/t_{\rm k} > 1\), indicating that the impulsive framework provides a reasonable description of the ejection process for these systems.

\begin{figure*}[ht!]
    \centering
   \includegraphics[width=1\linewidth]{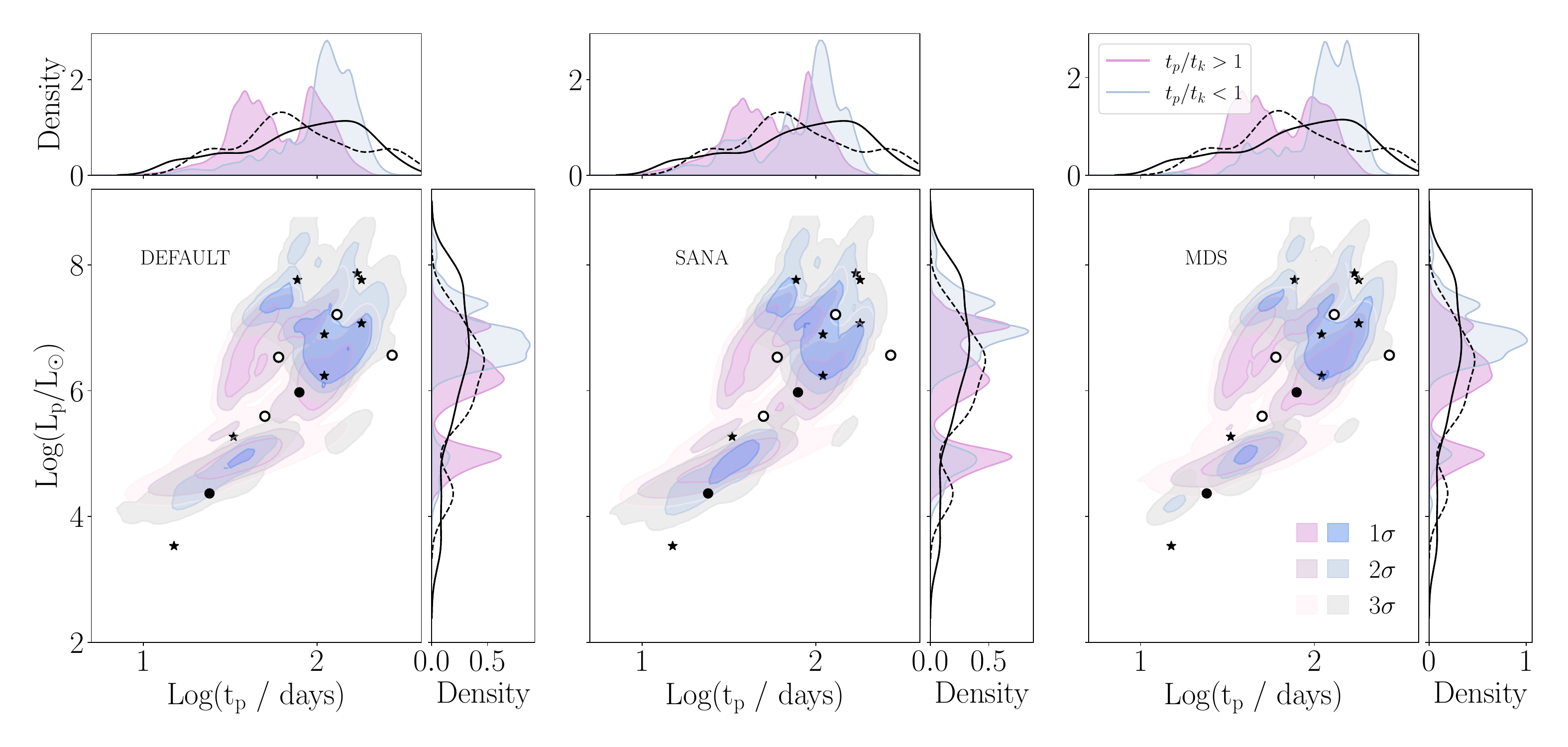}
    
    \caption{Comparison of the predicted plateau luminosity ($L_{90}$) and duration ($t_{90}$) distributions for Zone III systems with both the full observed LRN sample compiled by \citet{Matsumoto2022} (black solid line) and subset with pre-explosive images (black dashed line). Populations are divided by the ratio $t_{\rm p}/t_{\rm k}$: systems with $t_{\rm p}/t_{\rm k}<1$ (indicative of gradual mass ejection) are shown with filled \textit{blue} contours, and systems with $t_{\rm p}/t_{\rm k}>1$ with filled \textit{pink} contours. Contours indicate 1–3$\sigma$ density levels in increasing intensity. The figure highlights that systems with $t_{\rm p}/t_{\rm k}<1$ tend to occupy higher-luminosity, longer-duration regions, suggesting that the under-prediction of long-lived LRNe in impulsive-ejection models may be due to the instantaneous ejection assumption.}
    \label{fig:14}
\end{figure*}

The distribution of \(t_{\rm p}/t_{\rm k}\) depending on their CE outcomes is shown in the \textit{middle} panel of Figure~\ref{fig:13}. Notably, in the regime where \(t_{\rm p}/t_{\rm k} < 1\), the fraction of systems that undergo a stellar merger and a SCEE is similar. The \textit{right} panel shows the companion distribution for systems in Zone~III, indicating that donors and companions for which the impulsive approximation fails are primarily extended donors with WDs and MS companions that can disrupt during inspiral. If the inspiral occurs over multiple orbits, the envelope has time to respond dynamically, potentially altering the final outcome of the CE episode.

In such cases, an MS companion could, in principle, traverse less dense material and avoid disruption. This suggests that many more systems with MS companions might survive the first CE phase, expanding the population of systems that undergo a second CE across a broader \(M\)–\(R\) landscape than shown in Figure~\ref{fig:8} (\textit{top}). These extended evolutionary pathways link wide and close MS + WD binaries \citep[e.g.,][]{10.1093/mnras/stad4005} to distinct post-CE outcomes \citep[e.g.,][]{Vigna-Gomez2018,2021ApJ...920L..17V}, including the formation of double compact objects \citep[e.g.,][]{chattaraj2025formingdoubleneutronstars}, Thorne–\.Zytkow–like analogs \citep{2022MNRAS.513.4802A} via accretion-induced collapse, and calcium-rich supernovae that retain hydrogen envelopes \citep{Yadavalli2024}.

For systems with \(t_{\rm p}/t_{\rm k} < 1\) and relativistic companions, the inspiral can proceed over many orbits, allowing the envelope to respond dynamically and drive prolonged mass ejection. Any associated LRN-like transient would then be fainter and/or longer-lived. If the envelope is largely removed while the inspiral continuee, the compact object may ultimately merge with the donor core, resembling the delayed-merger scenario proposed by \citet{2022ApJ...932...84M} for LFBOTs and gamma-ray bursts \citep{2001ApJ...550..357Z,2024ApJ...977..196H}

Consequently, the observed LRN population is likely biased against systems in which the impulsive approximation breaks down, emphasizing the need for more sophisticated light-curve modeling and targeted observational strategies to uncover this hidden demographic of CE outcomes. These systems are also more likely to be dust-obscured \citep{2022ApJ...937...96M}, making them difficult to detect and thus underrepresented in observational samples despite potentially being common. This highlights the importance of accounting for survey selection effects when interpreting the observed demographics of CE events, as the detectable population may not accurately reflect the true underlying distribution.

Figure~\ref{fig:14} compares the predicted plateau luminosity and duration distributions for Zone~III systems with the observed LRN sample. The population is separated by the ratio \(t_{\rm p}/t_{\rm k}\): systems with \(t_{\rm p}/t_{\rm k} < 1\), indicative of gradual envelope ejection, are shown as filled \textit{blue} contours, while systems with \(t_{\rm p}/t_{\rm k} > 1\) are shown as filled \textit{pink} contours. The contours represent the 1–3\(\sigma\) density levels, with observed LRNe over-plotted for reference.

The \textit{blue} systems (\(t_{\rm p}/t_{\rm k} < 1\)) are poorly described by an impulsive ejection model; we consider adopting a non-impulsive treatment would shift the brightest of these systems toward longer plateau durations, bringing the predicted distribution into closer agreement with the observed LRNe. Conversely, systems in the low-luminosity, short-duration region are expected to become even dimmer under a gradual ejection scenario, likely placing them below current detection limits. Dust obscuration may further contribute to this observational bias, helping to explain why many observed LRNe do not lie in the densest regions of Zone~III, as noted in Figure~\ref{fig:2}.

As a result, the observed LRN population is likely biased toward mergers occurring near the boundary between Zones~II and~III, where transients are luminous enough to be detected. Systems with \(t_{\rm p}/t_{\rm k} < 1\) preferentially occupy the high-luminosity, long-timescale region, supporting the idea that under-prediction of long-lived LRNe arises from the impulsive ejection assumption, which produces brighter, shorter-lived transients. This further suggests that LRNe in the low-luminosity, short-duration region primarily correspond to mergers involving main-sequence or mildly evolved donors, while SCEE events in this regime may remain largely undetected.

\section{Summary and Conclusions}\label{sec:dis}
In this study, we combine rapid population synthesis with detailed stellar evolution models to characterize the mass, velocity, and launching radius of ejecta across a wide range of binary configurations, providing predictive insight into LRN properties. We map the landscape of COMPAS systems undergoing unstable mass transfer and CE evolution, focusing on the evolutionary states of both the donor (engulfing) and accretor (embedded) stars, while incorporating the effects of tidal disruption of the embedded companion. This approach directly connects pre-CE progenitor properties to the resulting transients and constrains the parameter space for SCEE. A comprehensive study of both successful and failed envelope ejections is essential for understanding how the CE phase shapes the evolutionary outcomes of binaries. By examining the full range of outcomes, we identify the key physical mechanisms and parameters that govern the luminosity function of LRNe. The principal findings of this work and their implications are:

\begin{itemize}
\item Unstable mass transfer and CE outcomes imprint a bimodal signature on the LRN luminosity function, sensitive to initial binary parameters and the donor’s evolutionary state (Section~\ref{sec_5.3}: Figure~\ref{fig:zones}). If confirmed by LSST, this bimodality, together with progenitor constraints, could allow the determination of LRN progenitors and outcomes based solely on light-curve properties.

\item A larger fraction of CE events result in mergers or partial ejections, primarily due to tidal disruption of the embedded MS companion (Table~\ref{tab:4}). In systems with pre-outburst imaging, this trend provides a diagnostic for distinguishing between WD and MS companions. The pronounced reduction in post-CE survivors, especially among binaries with MS companions, offers a clear, testable prediction for targeted observations of CE systems.

\item Most of the LRN luminosity function is reproduced, but the brightest events are underpredicted, indicating that additional energy sources beyond recombination are required (Figure~\ref{fig:hst_l90}). This highlights the need for improved modeling and targeted observations to understand the most luminous LRNe.

\item Highly extended progenitors should appear in observed samples under the impulsive ejection assumption. Their observed scarcity suggests that envelope ejection often proceeds gradually over multiple orbital timescales, producing fainter, longer-lived transients that are more difficult to detect (Section~\ref{sec_rethinking}, Figure~\ref{fig:14}). Accounting for this gradual ejection helps reconcile model predictions with observations and emphasizes the need for frameworks beyond the impulsive approximation.

\item CE interactions involving WD, NS, and BH accretors often result in SCEE outcomes, opening pathways to the formation of exotic and observationally intriguing systems. For example, red giants hosting embedded WDs have been proposed as analogs to Thorne–\.Zytkow objects \citep{2022MNRAS.513.4802A}, and as progenitors of calcium-rich supernovae that retain hydrogen envelopes. Merger outcomes in systems with NSs and BHs, may also provide a channel for powering LFBOTs under delayed inspiral conditions \citep{2022ApJ...932...84M}.

\end{itemize}

With the Vera C. Rubin Observatory and its Legacy Survey of Space and Time (LSST), the transient sky will be monitored with unprecedented cadence, dramatically expanding the discovery space for LRNe. For LRNe alone, the expected detection rate is set to increase from a handful of known events to hundreds per year \citep{Howitt2020}, transforming LRNe into a statistically rich population. This new era of big data in time-domain surveys offers the potential to unravel CE evolution, which is crucial for interpreting the transient Universe. The framework presented in this paper represents a step toward that goal. Early LSST detections of LRNe will provide compelling targets for follow-up observational campaigns, enabling a deeper understanding of how pre- and post-CE evolutionary pathways manifest. In this way, LRNe can evolve from a phenomenological classification into physical and dynamical probes of one of the least understood phases of binary evolution.

\section*{Acknowledgments}
Our perspectives on the topics discussed herein have been shaped  through insightful exchanges with R. Foley, I. Mandel, M. Gallegos-Garcia,  J. I. Cortes and T. Hutchinson-Smith. We acknowledge use of the lux supercomputer at UC Santa Cruz, funded by NSF MRI grant AST 1828315.  We acknowledge support by the Heising-Simons Foundation and the NSF (AST-2307710, AST-2206243, AST-1911206, and AST-1852393). The Villar Astro Time Lab acknowledges support through the David and Lucile Packard Foundation, National Science Foundation under AST-2433718, AST-2407922 and AST-2406110, as well as an Aramont Fellowship for Emerging Science Research. This work is supported by the National Science Foundation under Cooperative Agreement PHY-2019786 (The NSF AI Institute for Artificial Intelligence and Fundamental Interactions, http://iaifi.org/). 

\bibliography{bib}{}
\bibliographystyle{aasjournal}

\end{document}